\documentclass[aps,floatfix,onecolumn,a4paper,noshowpacs, nofootinbib,superscriptaddress,11pt]{revtex4}

\usepackage{booktabs}
\usepackage{multirow}
\usepackage{multirow,array}
\usepackage[utf8]{inputenc}
\usepackage{amsmath,amssymb,graphicx,bm}
\usepackage{geometry}
\usepackage{caption}
\usepackage{siunitx}
\usepackage{hyperref}
\hypersetup{colorlinks=true, linkcolor=blue, citecolor=blue, urlcolor=blue}
\usepackage{graphicx,float,tikz}
\usepackage[all]{xy}
\usepackage{bm,amsmath,upgreek,bigints}

\usepackage{amssymb}
\usepackage{color}
\usepackage{epsfig,bm}		
\usepackage{graphicx,epstopdf}
\usepackage{subfigure,upgreek}
\usepackage{pdfpages}
\usepackage{multirow}

\newcommand{\beq}{\begin{eqnarray}}
\newcommand{\eeq}{\end{eqnarray}}

\usepackage{bookmark,textgreek}
\usepackage{hyperref,color,xcolor}
\definecolor{lightgray}{gray}{0.95}
\hypersetup{hidelinks,colorlinks=true,breaklinks=true,urlcolor= blue}
\hypersetup{%
    colorlinks = true,
    linkcolor  = blue,
    citecolor = cyan,
  }
\usepackage{siunitx}
\definecolor{headergray}{gray}{0.90}

\usepackage[font=small,labelfont=bf,labelsep=space]{caption}
\usepackage{natbib}
\setcitestyle{square,numbers}

\newcommand\orcidroldao{{\href{https://orcid.org/0000-0003-3978-532X}{\orcidicon}}}

\newcommand\orcidcasadio{{\href{https://orcid.org/0000-0002-1330-7787}{\orcidicon}}}

\newcommand{\orcidicon}{%
	\begin{tikzpicture}
	\draw[lime, fill=lime] (0,0)
		circle [radius=0.16]
		node[white] {{\fontfamily{qag}\selectfont \tiny ID}};
	\draw[white, fill=white] (-0.0625,0.095)
		circle [radius=0.007];
	\end{tikzpicture}	\hspace{-2mm}
}

\begin{document}

\title{Black holes with axion hair from cubic Riemann interactions}

\author{R.~Casadio\orcidcasadio}
\email{casadio@bo.infn.it}
\affiliation{Dipartimento di Fisica e Astronomia, Universit\`a di Bologna, via Irnerio~46, 40126 Bologna, Italy}
\affiliation{I.N.F.N., Sezione di Bologna, I.S.~FLAG, viale B.~Pichat~6/2, 40127 Bologna, Italy}
\affiliation{Alma Mater Research Center on Applied Mathematics (AM\,$^2$), Via Saragozza 8, 40123 Bologna, Italy}

\author{R.~da~Rocha\orcidroldao}
\email{roldao.rocha@ufabc.edu.br}
\affiliation{Federal University of ABC, Center of Mathematics, Santo Andr\'e, S\~ao Paulo, 09580-210, Brazil}
\bigskip
\begin{abstract}
An effective field theory (EFT) of gravity with a cubic Riemann invariant coupled to an axion-like field
is investigated and shown to propagate no additional independent low-energy degrees of freedom.
This interaction generates a curvature-induced scalar profile around black holes,
characterised by a secondary axion charge fixed by regularity of the axion field at the horizon,
while the requirement of a regular event horizon constrains the asymptotic axion background.
We construct the corresponding black hole solutions and show that the backreaction of the axion field
could affect the metric only at subleading order in the EFT expansion.
The modified geometry impacts black hole thermodynamics and Hawking evaporation,
thus providing a controlled realisation of black holes with higher-curvature contributions and secondary axion
hair motivated by ultraviolet physics.
The modified Hawking temperature allows for the existence of black hole remnants at the end
of the evaporation within the regime of validity of the EFT. 
\end{abstract}

\maketitle

\section{Introduction}

Axion-like pseudoscalar fields arise naturally as pseudo-Nambu--Goldstone bosons associated with spontaneously broken global symmetries, with the QCD axion providing the canonical realization through the Peccei--Quinn (PQ) mechanism~\cite{Peccei:1977ur,Wilczek:1977pj,Weinberg:1977ma}.
At low energies, axions couple weakly to gauge fields through higher-dimensional operators, most notably via the axion-photon interaction, enabling axion-photon conversion in external electromagnetic backgrounds. In the early universe, post-inflationary breaking of the PQ symmetry can lead to the formation of overdense regions, giving rise to axion miniclusters~\cite{Hogan:1988mp,Casadio:2023mgl}.
Subsequent nonlinear evolution, driven by gravitational interactions and dissipative processes, can produce self-gravitating configurations known as axion stars~\cite{Levkov:2018kau,Bai:2021nrs}.
In sufficiently dense regimes, axion self-interactions may become effectively attractive, leading to coherent configurations and subsequent gravitational collapse~\cite{Barranco:2012ur,deFreitasPacheco:2018gpi,Chen:2023vet}.
These considerations motivate the study of axion dynamics in strong-gravity environments, where spacetime curvature can substantially modify the scalar dynamics~\cite{Gorghetto:2024vnp}. 

Higher-dimensional operators coupling axions to curvature invariants arise naturally upon integrating out heavy degrees of freedom at a cutoff scale, in the spirit of ultraviolet (UV) completions of
gravity~\cite{Gross:1986iv,Metsaev:1987zx,Svrcek:2006yi,Donoghue:2017vvl,Burgess:2003jk}.
Curvature invariants appear as higher-order corrections to the Einstein--Hilbert action. While quadratic curvature corrections have been extensively studied, cubic contractions of the Riemann tensor constitute the genuinely new leading gravitational EFT corrections beyond quadratic order, corresponding to the unique local counterterm that appears in the two-loop effective action of pure
gravity~\cite{Goroff:1985th,vandeVen:1991gw,Donoghue:1993eb,Bueno:2023jtc}.
Among these, cubic gravity provides a higher-curvature extension of general relativity (GR) in which UV-sensitive corrections modify the nonlinear dynamics while preserving the low-energy propagating
spectrum~\cite{DeFelice:2023vmj,Lessa:2023thi,Rachwal:2021bgb,Kulkarni:2024ghc,daRocha:2024lev}.
In this context, quantum gravitational hair was explored in Refs.~\cite{Calmet:2021lny,Calmet:2019eof,daRocha:2021sqd,Perrucci:2024qrr,Calmet:2021stu,daRocha:2020gee,Abdalla:2009pg}.
Related quantum and higher-curvature modifications of black hole geometries, together with their consequences for near-horizon physics and associated observables, have also been investigated in
Refs.~\cite{Battista:2023iyu,Wang:2025fmz,
Tello-Ortiz:2024mqg,daRocha:2012pt,Panotopoulos:2025ygq,Cavalcanti:2016mbe,Buoninfante:2025neu}.

The simultaneous presence of a light axion-like degree of freedom and higher-curvature gravitational operators raises a further question
that is absent in pure cubic gravity, namely whether the Wilson coefficients controlling the higher-curvature sector could depend on
dynamical low-energy fields.
In this case, the coefficient of the cubic curvature invariant is promoted from a fixed EFT parameter to a
field-dependent quantity.
Strong spacetime curvature can then source the axion sector,~\footnote{We will show that the cubic Riemann invariant
acts as an inhomogeneous source in the axion equation of motion. Consequently, a black hole can induce a
nontrivial axion profile and no independent scalar charge is introduced.}
producing a curvature-induced scalar configuration
which subsequently backreacts on the geometry. This provides a
dynamical mechanism for generating secondary hair which does not appear in pure cubic
gravity, where the corresponding Wilson coefficient is fixed~\cite{Bueno:2016ypa,Bueno:2018xqc}.
Black holes are then the natural candidates to investigate this mechanism,
since these vacuum geometries possess nonvanishing Riemann invariants
capable of sourcing the axion even in the absence of an independent
scalar charge.

The coupling between an axion field and a cubic Riemann invariant provides a symmetry-consistent
higher-derivative interaction beyond the minimal QCD axion sector. Unlike familiar scalarization mechanisms,
including Gauss--Bonnet, Chern--Simons, and Horndeski
models~\cite{Sotiriou:2015pka,Sotiriou:2013qea,Alexander:2008wi,Kobayashi:2019hrl,Doneva:2017bvd,Gonzalez-Espinoza:2026gen},
the cubic interaction considered here does not generate an independent primary scalar charge,
but instead induces a curvature-driven secondary hair axion scalar profile.
It induces corrections to near-horizon observables and black hole thermodynamics,
providing a controlled probe of higher-curvature effects.
As demonstrated in Appendix~\ref{sec2}, no additional propagating degrees of freedom
arise within this regime, and the quadratic action for fluctuations remains free of pathological instabilities. 
We obtain the modified metric explicitly and analyse its properties.
In particular, we show that the backreaction of axion-induced corrections on the metric remains
subleading within the regime of the EFT employed here.
We then investigate the implications for black hole thermodynamics using the Iyer--Wald formalism~\cite{Iyer:1994ys}
and derive corrections to the Hawking temperature~\cite{Hawking:1975vcx}, Wald entropy, heat capacity, and free energy.

This paper is organized as follows. In Sec.~\ref{sec1}, we formulate the EFT with an axion-like field linearly
coupled to the cubic Riemann invariant and construct the resulting curvature-sourced scalar profile.
We also obtain the metric solutions carrying the interaction between the axion field and the cubic curvature, determine the event horizon, and derive the regularity condition that constrains the asymptotic axion background.
In Sec.~\ref{sec:bh_chemistry}, we analyse the implications of this EFT for black hole thermodynamics
and semiclassical evaporation.
Using the Iyer--Wald formalism, we compute the corrections to the Wald entropy, Hawking temperature,
Hawking emission rate, and evaporation time, induced by higher-curvature interactions and their axion backreaction. The modified Hawking temperature allows evaporation to end in a black hole remnant, within the perturbative EFT regime.
Sec.~\ref{secfim} summarises our results and outlines relevant extensions.
Appendix~\ref{sec2} investigates  the perturbative structure of the theory, demonstrating the absence of
additional propagating degrees of freedom within the EFT regime and the regularity of the quadratic action for fluctuations. 
Appendix~\ref{appb1} analyses the axion-cubic-curvature corrections in terms of an effective anisotropic matter
sector and derives the corresponding energy density and radial and tangential pressures. 
Appendix~\ref{secPPN} examines the weak-field phenomenology of the solution,
deriving its post-Newtonian expansion and identifying the leading quadratic correction through a theory-specific
2PN parameter.
Solar-System tests are then compared with the strong-field photon-sphere scale and with the perturbative domain
of the EFT, thereby connecting weak- and strong-field probes of the cubic-curvature interaction.

\section{EFT and black hole solutions} % in cubic curvature gravity with ultrasoft secondary hair}
\label{sec1}

The EFT employed in this work is organised as a derivative expansion about GR,
where higher-curvature operators encode the leading UV corrections to the Einstein--Hilbert
action~\cite{Vacaru:2024dvc,Matyjasek:2020bzc,Marciu:2020ski}.
Among such operators, the cubic Riemann contraction
$
\mathcal{O}_{RRR}
=
R_{\mu\nu\alpha\beta}
R^{\alpha\beta}{}_{\rho\sigma}
R^{\rho\sigma\mu\nu}
$ 
plays a distinguished role. In four spacetime dimensions, pure Einstein gravity is finite on shell at one loop~\cite{tHooft:1974toh}.
The first genuine UV divergence appears at two loops and is proportional to the relevant nontrivial cubic curvature
invariant~\cite{Goroff:1985th}.
Consequently, $\mathcal{O}_{RRR}$ is the leading purely gravitational local operator that encodes the two-loop structure
of quantum gravity in the EFT expansion.
%The $\mathcal{O}_{RRR}$ operator provides the natural curvature invariant governing the leading EFT interaction studied here. 
Phenomenological constraints on cubic-curvature EFTs have also been discussed in
Refs.~\cite{daRocha:2021xwq,daRocha:2023waq,Marciu:2023hdb}.

In consistent UV completions, higher-curvature interactions arise systematically in the derivative expansion of the effective
gravitational action~\cite{Agrawal:2024ejr}, with cubic curvature invariants appearing at higher order in the expansion.
In many UV-complete frameworks, pseudoscalar degrees of freedom can couple linearly to higher-curvature invariants.
The corresponding low-energy effective action can then be written, in units with $\hbar=c=1$, as 
\begin{equation}\label{act1}
S = \int d^4x \,\sqrt{-g} \left(
\frac{R}{16\pi G}
-\frac{1}{2} \nabla_\mu \mathfrak a \nabla^\mu \mathfrak a
+ \frac{c_6\, G}{6\,f} \mathfrak a\, R_{\mu\nu\alpha\beta} R^{\alpha\beta}{}_{\rho\sigma} R^{\rho\sigma\mu\nu}
\right)
\ ,
\end{equation}
where $R$ is the Ricci scalar of the Riemann tensor $R_{\mu\nu\alpha\beta}$ and $G=M_{\rm Pl}^{-2}$ is Newton's constant.
Here $\mathfrak a = \mathfrak a(x^\mu)$ denotes the pseudoscalar axion field and $c_6$ the dimensionless Wilson
coefficient encoding the UV-sensitive strength of the axion-dressed cubic-curvature operator.
This coupling is suppressed by the axion decay constant $f$, associated with the spontaneous breaking of the PQ
symmetry \cite{Cicoli:2026fqp}, which sets the corresponding dimensionful EFT coupling 
\begin{equation}
\alpha_6
\equiv
\frac{c_6}{f}
\ .
\label{A6}
\end{equation}
Ref.~\cite{Giddings:1987cg} showed that the Kalb--Ramond 2-form field $B_{\mu\nu}$, with strength $H = dB$,
is the axion Hodge dual $H_{\mu\nu\rho} = \epsilon_{\mu\nu\rho\sigma}\,\partial^\sigma \mathfrak{a}$.
Therefore the axion kinetic term in~\eqref{act1} takes the usual form
$\nabla_\mu \mathfrak{a} \, \nabla^\mu \mathfrak{a}=\frac{1}{6}\, H_{\mu\nu\rho} H^{\mu\nu\rho}$. 
The EFT remains perturbatively under control when the characteristic curvature scale lies well below
the Planck scale.
For a Schwarzschild black hole, the near-horizon curvature scales as $R\sim(GM)^{-2}$,
so that the corresponding dimensionless expansion parameter behaves as $G\, R\sim(GM^2)^{-1}$.
Consequently, the semiclassical regime $G\,M^2\gg 1$ ensures that higher-curvature corrections remain
perturbatively suppressed in the black hole exterior.

The coupling in Eq.~\eqref{act1} is well motivated from the EFT perspective, since the
same invariant $\mathcal{O}_{RRR}$ arises as the two-loop counterterm of pure Einstein gravity in the
Goroff--Sagnotti calculation~\cite{Goroff:1985th}.
In our work, this operator is dressed by the dynamical factor $\mathfrak a/f$ and treated perturbatively
rather than as a fundamental modification of the gravitational spectrum.
Although higher-curvature interactions may generically introduce additional modes,
Appendix~\ref{sec2} shows that the cubic interaction generates only EFT-suppressed corrections
to the Einstein--Hilbert kinetic operator within the regime of validity of the derivative expansion.
The spin-2 sector retains the massless graviton pole continuously connected to GR,
while additional roots occur only at scales where the EFT expansion breaks down.
Thus, the axion-dressed Goroff--Sagnotti operator introduces no propagating ghost
within the domain of validity of the EFT.

Black hole solutions will be constructed perturbatively around the Schwarzschild background,
with the expansion organised simultaneously in powers of spacetime curvature and in the EFT
parameter 
\begin{equation}
\epsilon_{\rm EFT}\equiv \frac{\alpha_6}{G^2 M^3} \ll 1
\ .
\label{Eeft}
\end{equation}
The metric deformation induced directly by the cubic invariant appears at order $\mathcal{O}(\alpha_6)$,
whereas the axion stress tensor contributes only at order $\mathcal{O}(\alpha_6^2)$.
Varying the action~\eqref{act1} with respect to $\mathfrak{a}$ yields the sourced wave equation governing
the axion field
\begin{equation}\label{swe1}
\Box \mathfrak{a}
=
-\frac{\alpha_6G}{6}
R_{\mu\nu\alpha\beta}
R^{\alpha\beta}{}_{\rho\sigma}
R^{\rho\sigma\mu\nu}
\ .
\end{equation}
Unlike scalarization scenarios based on tachyonic instabilities, the axion is here linearly and uniquely sourced
by the curvature invariant.
The solution is fully fixed by the background geometry and boundary conditions and does not involve any
independent scalar charge or multiple solution branches~\cite{Sotiriou:2015pka,Ballesteros:2023iqb}. 

The axion-cubic curvature interaction induces black hole hair. Evaluating the cubic invariant on the
Schwarzschild background,
\begin{equation}
\label{ans1}
ds^2
=
-B(r)\,dt^2
+\frac{dr^2}{A(r)}
+r^2\,d\Omega^2
\ ,
\end{equation}
with $A=B=1-2GM/r$, yields
\begin{equation}
\label{1152}
R_{\mu\nu\alpha\beta} R^{\alpha\beta}{}_{\rho\sigma} R^{\rho\sigma\mu\nu}
=
96\frac{G^3 M^3}{r^9}
\ .
\end{equation}
All curvature invariants and wave operators are evaluated on the background Schwarzschild geometry,
consistently with the EFT expansion. 
Therefore, Eq.~(\ref{swe1}) reduces to
\begin{equation}\label{ino1}
\Box \mathfrak{a}
=
-16\,\alpha_6\,\frac{G^4 M^3}{r^9}
\ ,
\end{equation}
which can be solved using the static Green’s function of the wave operator.
One therefore obtains the purely curvature-induced axion profile
\begin{equation}\label{ainf}
\mathfrak{a}(r)
=
\mathfrak{a}_{\infty}
+\alpha_6
\left(
\frac{1}{24G^2M^3r}
+\frac{1}{24GM^2r^2}
+\frac{1}{18Mr^3}
+\frac{G}{12r^4}
+\frac{2G^2M}{15r^5}
+\frac{2G^3M^2}{9r^6}
\right)
+
\mathcal O(\alpha_6^2)
\ ,
\end{equation}
where $\mathfrak{a}_\infty$ is the asymptotic value of the axion field
at spatial infinity.
We remark that Eq.~(\ref{ainf}) is the exact radial solution at $\mathcal O(\alpha_6)$
that is also regular on the horizon $r\simeq 2\,G\,M$.
Its asymptotic behaviour can be written as
\begin{equation}
\mathfrak a(r)
=
\mathfrak a_\infty
+\frac{Q_{\mathfrak a}}{r}
+\mathcal O(r^{-2}),
\end{equation}
with the scalar charge 
\begin{equation}\label{scc1}
Q_{\mathfrak a}
=
\frac{\alpha_6}{24G^2M^3}
\ ,
\end{equation}
which  is uniquely fixed by the curvature source and horizon regularity,
and therefore constitutes a secondary scalar charge rather than primary scalar hair.
The axion profile~\eqref{ainf} can thus be interpreted as a curvature-induced
polarisation cloud whose radial structure is fixed by the black hole mass
and EFT coupling.
In particular, the axion configuration remains finite and can become non-negligible
in the strong-curvature regime near the black hole.
It is worth emphasising that no explicit $\mathcal O(\alpha_6^2)$ correction to the axion profile is
required for the present analysis, since the leading axion solution
already generates the scalar stress tensor and scalar-dependent
cubic-curvature contribution at $\mathcal O(\alpha_6^2)$.

Varying the cubic interaction term in the action (\ref{act1}) with respect to the metric generates
both purely geometric higher-curvature contributions and terms involving derivatives of the axion field.
The latter arise from the metric dependence of the Riemann tensor and, after integration by parts,
lead to couplings between curvature and derivatives of $\mathfrak{a}$, which, using the axion profile~\eqref{ainf},
read  
\begin{equation}\label{cont1}
\alpha_6\, G\,
R_{\mu\alpha\beta\gamma}
R_{\nu}{}^{\alpha\beta\gamma}
\,\nabla^\mu\nabla^\nu \mathfrak{a}
=
-192\,\alpha_6^2\,
\frac{G^7 M^5}{r^{15}}
\ ,
\end{equation}
where we included the factor $\alpha_6\, G$ from the coupling in Eq.~\eqref{act1}.
The perturbative hierarchy can be made explicit by expanding the metric and axion field
around the Schwarzschild background, up to $\mathcal O(\alpha_6^3)$, as
\begin{subequations}
\begin{eqnarray}
g_{\mu\nu}
&\simeq&
g_{\mu\nu}^{(0)}
+
\alpha_6\, h_{\mu\nu}^{(1)}
+
\alpha_6^2\, h_{\mu\nu}^{(2)}
\ ,
 \\
\mathfrak a
&\simeq&
\mathfrak a_\infty
+
\alpha_6\,\mathfrak a_{(1)}
+
\alpha_6^2\,\mathfrak a_{(2)}
\ ,
\end{eqnarray}
\end{subequations}
where $\alpha_6\,\mathfrak a_{(1)}$ is the leading order curvature-induced
profile given in Eq.~\eqref{ainf}.
The cubic-curvature interaction contains at leading order a term
proportional to $\alpha_6\,\mathfrak a_\infty$, which contributes to the
metric equations at $\mathcal O(\alpha_6)$ for $\mathfrak a_\infty\neq0$.
Contributions involving the curvature-induced profile $\mathfrak a_{(1)}$ carry an additional
power of $\alpha_6$ and are therefore $\mathcal O(\alpha_6^2)$.
At the same order, the axion equation receives corrections from evaluating both the wave operator and the
cubic curvature invariant on the perturbed geometry $g_{\mu\nu}^{(0)}+\alpha_6 h_{\mu\nu}^{(1)}$.
Thus, $\mathfrak a_{(2)}$ is sourced by the first-order metric deformation
and fixed by the corresponding second-order axion equation subject to
the same asymptotic and horizon-regularity conditions.
Since $h_{\mu\nu}^{(1)}$ is asymptotically suppressed, $\alpha_6^2\,\mathfrak a_{(2)}$
remains subleading at spatial infinity.
The leading-order profile in Eq.~\eqref{ainf}) is therefore sufficient to
determine the axion-mediated metric backreaction consistently through $\mathcal O(\alpha_6^2)$.

The axion stress tensor to $\mathcal O(\alpha_6^2)$ reads
\begin{eqnarray}\label{ax1}
\!\!T_{\mu\nu}^{(\mathfrak a)}
&\!=\!&
\nabla_\mu \mathfrak a \, \nabla_\nu \mathfrak a
-\frac{1}{2}g_{\mu\nu}
g^{\rho\sigma}
\nabla_\rho \mathfrak a \, \nabla_\sigma \mathfrak a
\nonumber
\\
&\!\simeq\!&
\frac{\alpha_6^2}{576 G^4 M^6 r^4}\!
\left(
1\!+\!\frac{2GM}{r}
\!+\!\frac{4G^2M^2}{r^2}
\!+\!\frac{8G^3M^3}{r^3}
\!+\!\frac{16G^4M^4}{r^4}
\!+\!\frac{32G^5M^5}{r^5}
\right)^2
\!\!A_{\mu\nu}
\ ,
\end{eqnarray}
where $\displaystyle A_{\mu\nu}=\delta_\mu^{\,r}\delta_\nu^{\,r}-\frac{1}{2}g_{\mu\nu}g^{rr}$.
The corresponding energy density is given by 
\begin{eqnarray}
\!\!\!\!\!\!\!\!\rho_{\mathfrak a}\!
&\!=\!&
\!-T^{(\mathfrak a)t}{}_{t}
=
\frac{1}{2}g^{rr}(\partial_r \mathfrak a)^2
\nonumber
\\
\!\!\!\!\!\!
&\!\simeq\!&
\!\frac{\alpha_6^2}{1152 G^4 M^6 r^4}\!
\left(1\!-\!\frac{2GM}{r}\!\right)\!
\left(
1\!+\!\frac{2GM}{r}
\!+\!\frac{4G^2M^2}{r^2}
\!+\!\frac{8G^3M^3}{r^3}
\!+\!\frac{16G^4M^4}{r^4}
\!+\!\frac{32G^5M^5}{r^5}
\right)^2
\ ,
\end{eqnarray}
whereas the radial and tangential pressures are, respectively, given by
\begin{eqnarray}\label{prpt}
p_{\mathfrak a,r}
&=&
T^{(\mathfrak a)r}{}_{r}
=
\frac{1}{2}g^{rr}(\partial_r \mathfrak a)^2
=
\rho_{\mathfrak a}
\ ,
\nonumber\\
p_{\mathfrak a,\perp}
&=&
T^{(\mathfrak a)\theta}{}_{\theta}
=
T^{(\mathfrak a)\phi}{}_{\phi}
=
-\frac{1}{2}g^{rr}(\partial_r \mathfrak a)^2
=
-\rho_{\mathfrak a}
\ .
\end{eqnarray}
The axion sector therefore exhibits the anisotropic stress structure characteristic of a purely radial scalar gradient.
The full effective energy-momentum in the modified Einstein equations also receives independent contributions
from the metric variation of the cubic curvature interaction.
Restricting attention to the axion sector~\eqref{prpt} therefore implies the non-vanishing pressure anisotropy 
$p_{\mathfrak a,r}-p_{\mathfrak a,\perp}=2\rho_{\mathfrak a}$.

To determine the metric corrections generated by the axion-cubic-curvature
interaction, we consider the static and spherically symmetric ansatz~\eqref{ans1} with
$A(r)$ and $B(r)$ to determine.
It is then useful to expand the metric functions in powers of $\alpha_6$ and write
\begin{eqnarray}
\label{bm1}
B(r)
&\simeq&
B_0(r)+\alpha_6\,B_{(1)}(r)
+\alpha_6^2\,B_{(2)}(r),\\
A(r)
&\simeq&
A_0(r)+\alpha_6\,A_{(1)}(r)
+\alpha_6^2\,A_{(2)}(r)
\ ,
\label{am1}
\end{eqnarray}
where
\begin{equation}
B_0(r)=A_0(r)=1-\frac{2GM}{r}
\end{equation}
is the Schwarzschild background. 

The field equations stemming from the action~\eqref{act1} can be written as
\begin{equation}
\label{feq1}
G_{\mu\nu}
=
8\pi G
\left[
T_{\mu\nu}^{(\mathfrak a)}
+
T_{\mu\nu}^{(R^3)}
\right]
\ ,
\end{equation}
where $T_{\mu\nu}^{(\mathfrak a)}$  is given in Eq.~\eqref{ax1} and the effective curvature contribution
is defined directly by the metric
variation
\begin{equation}\label{tr3}
T_{\mu\nu}^{(R^3)}
\equiv
-\frac{2}{\sqrt{-g}}\,
\frac{\delta}{\delta g^{\mu\nu}}
\left(
\sqrt{-g}\,
\frac{\alpha_6 G}{6}\,
\mathfrak a\,
R_{\rho\sigma\alpha\beta}
R^{\alpha\beta}{}_{\gamma\delta}
R^{\gamma\delta\rho\sigma}
\right)
\ .
\end{equation}
After fixing the asymptotic normalisation of the time coordinate and keeping the ADM~\cite{Arnowitt:1959ah}
mass $M$ unchanged, one obtains
\begin{align}\label{a1}
B_{(1)}(r)
&=
\frac{320\pi}{3}\,
\mathfrak a_\infty
\frac{G^5M^3}{r^7},
\\
A_{(1)}(r)
&=
576\pi\,\mathfrak a_\infty
\frac{G^4M^2}{r^6}
\left(
1-\frac{49GM}{27r}
\right)
\ .
\end{align}
Thus, the linear order deformation of the metric is proportional to the asymptotic axion
background $\mathfrak a_\infty$.
At this order, the deformation originates from the variation of the cubic-curvature interaction
evaluated on the constant asymptotic axion background, whereas the dynamical axion stress tensor
starts contributing only at $\mathcal O(\alpha_6^2)$.

At second order, the field equations contain the iteration of the
first-order geometric deformation, the metric variation of the
cubic-curvature interaction involving the curvature-induced axion
profile, and the axion stress-energy backreaction.
Solving the complete second-order equations directly, with the same asymptotic time
normalisation and ADM mass conditions, yields the modified metric functions
\begin{eqnarray}
B(r)
&\simeq&
1-\frac{2GM}{r}
+\frac{320\pi\,\mathfrak a_\infty \alpha_6 G^5M^3}{3r^7}
\nonumber\\
&&
+{\alpha_6^2\,\pi}
\left[
\frac{1}{432G^2M^5r^3}
+\frac{1}{216GM^4r^4}
+\frac{1}{120M^3r^5}
+\frac{2 G}{135M^2r^6}+\frac{11\pi G^2}{27Mr^7}
+\frac{97 G^3}{21r^8}
\right.
\nonumber
\\
&&
\left.\qquad
\quad+\frac{383 G^4M}{81r^9}
+\frac{32 G^5M^2}{5r^{10}}
+\frac{64 G^6M^3}{33r^{11}}
\left(
5+110592\pi G\mathfrak a_\infty^2
\right)
\right.
\nonumber
\\
&&
\quad\left.\qquad
+\frac{32 G^7M^4}{297r^{12}}
\left(
145\!-\!8273664\pi G\mathfrak a_\infty^2
\right)
\!+\!\frac{32 G^8M^5}{27r^{13}}
\left(
11\!+\!765792\pi G\mathfrak a_\infty^2
\right)
\right]
\ ,
\label{br1}
\\
\!\!\!\!\!\!\!\!\!\!\!
A(r)
&\simeq&
1-\frac{2GM}{r}
+
576\pi\,\mathfrak a_\infty
\alpha_6
\frac{G^4M^2}{r^6}
\left(
1-\frac{49GM}{27r}
\right)
\nonumber\\
&&
+{\alpha_6^2\,\pi}
\left[
\frac{1}{144G^3M^6r^2}
+\frac{1}{144G^2M^5r^3}
+\frac{1}{108GM^4r^4}
+\frac{1}{72M^3r^5}+\frac{\pi G}{45M^2r^6}\right.
\nonumber\\
&&\quad\left.\qquad
+\frac{793 G^2}{27Mr^7}-\frac{1228 G^3}{63r^8}
-\frac{103 G^4M}{9r^9}
-\frac{800 G^5M^2}{81r^{10}}
-\frac{448 G^6M^3}{45r^{11}}\right.
\nonumber\\
&&\quad\left.\qquad
+\frac{64 G^7M^4}{495r^{12}}
\left(
984960\pi G\mathfrak a_\infty^2-83
\right)
-\frac{32 G^8M^5}{9r^{13}}
\left(
161+59232\pi G\mathfrak a_\infty^2
\right)
\right]
\ ,
\label{ar1}
\end{eqnarray}
where $\mathfrak a_\infty$ is yet to be determined.

The condition $g^{rr}(r_H^{(A)})=A(r_H^{(A)})=0$ determines a marginally trapped surface.
Independently, the vanishing of the norm of the static Killing vector,
$g_{tt}(r_H^{(B)})=-B(r_H^{(B)})=-0$, determines the corresponding candidate Killing
horizon.
A regular Schwarzschild-like black hole horizon therefore requires these
two surfaces to coincide order by order in the EFT expansion.
Solving the two conditions independently, at $\mathcal O(\alpha_6^2)$, yields 
\begin{eqnarray}
r_H^{(B)}
&\simeq&
2GM
-\frac{5\pi\mathfrak a_\infty\alpha_6}{3GM^3}
-\frac{\pi\alpha_6^2
\left(14400\pi G\mathfrak a_\infty^2+1\right)}
{1728\,G^4M^7}
\ ,
\label{rHA}
\\
r_H^{(A)}
&\simeq&
2GM
-\frac{5\pi\mathfrak a_\infty\alpha_6}{3GM^3}
+\frac{\pi\alpha_6^2
\left(1920\pi G\mathfrak a_\infty^2-1\right)}
{288\,G^4M^7}
\ .
\label{rHB}
\end{eqnarray}
The two roots coincide at linear order in $\alpha_6$ for any values of $\mathfrak a_\infty$,
whereas at quadratic order their difference is given by the expression 
\begin{equation}
r_H^{(B)}-r_H^{(A)}
\simeq
\frac{5\pi\alpha_6^2}{1728\,G^4M^7}
\left(1-5184\pi G\mathfrak a_\infty^2\right)
\ .
\end{equation}
Hence, requiring a common regular horizon at this EFT order fixes
\begin{equation}
\mathfrak a_\infty
=
\pm\frac{1}{72\sqrt{\pi G}}
\ .
\label{ainf-horizon}
\end{equation}
Choosing the positive value in Eq.~\eqref{ainf-horizon} and taking into account
Eq.~\eqref{ainf-horizon}, one then finds that the event horizon $r_H^{(B)}=r_H^{(A)}$
is located at
\begin{equation}
r_H
\simeq
2GM
-\frac{5\sqrt{\pi}\,\alpha_6}{216\,G^{3/2}M^3}
-\frac{17\pi\alpha_6^2}{7776\,G^4M^7}
\ .
\label{rh1}
\end{equation}
This result shows that the deviation from the Schwarzschild horizon is strongly suppressed for
large black holes and becomes progressively more significant toward smaller masses,
precisely where higher-curvature effects are expected to become more relevant.
It is also worth emphasising that both terms in Eq.~\eqref{rh1} shift the event horizon inward,
relative to the Schwarzschild value, for $\alpha_6>0$.
Since the linear corrections in Eqs.~\eqref{rHA} and~\eqref{rHB} are proportional to
$\mathfrak a_\infty\,\alpha_6$, the negative value of $\mathfrak a_\infty$ in Eq.~\eqref{ainf-horizon}
will give the same results for $\alpha_6<0$.

It is important to remark that we are here considering only vacuum solutions and the
values~\eqref{ainf-horizon} are fixed by horizon regularity, which would not be relevant
for a regular astrophysical source, like a star.
In that case, the constant $\mathfrak a_\infty$ should instead be fixed by matching conditions at the
star surface, which requires solving the field equations inside the matter source.
This task is not considered here.

The leading large-$r$ behaviour of the metric functions (\ref{br1}, \ref{ar1}) is  respectively expressed as
\begin{eqnarray}\label{bbbr1}
B(r)
&\simeq&
1-\frac{2GM}{r}
+
\frac{\pi \alpha_6^2}{432G^2M^5r^3}
\ ,
\\
A(r)\label{aaar1}
&\simeq&
1-\frac{2GM}{r}
+
\frac{\pi \alpha_6^2}{144G^3M^6r^2}
\ .
\end{eqnarray}
Although the leading corrections now scale as $r^{-3}$ in (\ref{bbbr1}) and $r^{-2}$ in (\ref{aaar1}),
no new $\mathcal O(r^{-1})$ contribution is generated, and the ADM mass therefore remains unchanged.
In particular, the leading-order expressions (\ref{bbbr1}) and (\ref{aaar1}) remain valid in the case of a star.

Taking into account both the canonical axion stress tensor~\eqref{ax1} and the effective
contribution generated by the metric variation of the cubic interaction~\eqref{tr3},
the total effective energy density and pressures associated with the corrected geometry
can be defined directly through the Einstein tensor as
$
\rho_{\rm eff}
=
-{G^t{}_t}/{8\pi G}, \;
p_r^{\rm eff}
=
{G^r{}_r}/{8\pi G}$, and  
$p_\perp^{\rm eff}=
{G^\theta{}_\theta}/{8\pi G}.$ 
Replacing the metric functions~\eqref{br1} and \eqref{ar1} provides a direct geometric characterisation
of the complete effective source, without requiring a separate splitting of the higher-curvature contribution.
Therefore, using the horizon-regular scalar profile and the corresponding metric
coefficients, the effective energy density, the radial pressure, and the tangential pressure are calculated
and respectively given by Eqs.~\eqref{rhoeff1}-\eqref{pteff1}. 
Although the linear contribution starts at $\mathcal O(r^{-8})$, the horizon-regular curvature-induced
axion profile generates at $\mathcal O(\alpha_6^2)$ a more slowly decaying $\mathcal O(r^{-4})$ contribution.
Consequently, at fixed nonzero $\alpha_6$, the leading large-$r$ behaviour of the complete effective source
can be expressed as 
\begin{equation}\label{effleading}
\rho_{\rm eff}
=
p_r^{\rm eff}
=
-p_\perp^{\rm eff}
=
\frac{\alpha_6^2}{1152G^4M^6r^4}
+\mathcal O(r^{-5})
\ .
\end{equation}
This asymptotic structure coincides with that generated by the
radial gradient of the horizon-regular axion configuration. 

Defining the radial and tangential equation-of-state parameters,
respectively, as
$
w_r(r)
=
{p_r^{\rm eff}(r)}/{\rho_{\rm eff}(r)}
$
and
$
w_\perp(r)
=
{p_\perp^{\rm eff}(r)}/{\rho_{\rm eff}(r)}
$,
and the effective mean parameter
\begin{equation}
w_{\rm eff}(r)
\equiv
\frac{p_r^{\rm eff}(r)+2p_\perp^{\rm eff}(r)}
{3\rho_{\rm eff}(r)}
\ ,
\end{equation}
one finds, in the asymptotic region,
\begin{subequations}\label{wasym}
\begin{eqnarray}
\lim_{r\to\infty}w_r(r)&=&1
\ ,\label{wa1}\\
\lim_{r\to\infty}w_\perp(r)&=&-1
\ ,\label{wa2}\\
\lim_{r\to\infty}w_{\rm eff}(r)&=&-\frac13
\ .\label{wa3}
\end{eqnarray}
\end{subequations}
 The pressure anisotropy,
\begin{equation}
\Delta_{\rm eff}(r)
\equiv
p_\perp^{\rm eff}(r)-p_r^{\rm eff}(r)
\ ,
\label{wa4}
\end{equation}
accordingly satisfies
\begin{equation}
\lim_{r\to\infty}
\frac{\Delta_{\rm eff}(r)}{\rho_{\rm eff}(r)}
=-2
\ .
\end{equation}
The nonvanishing asymptotic anisotropy shows that the effective source
does not approach an isotropic perfect-fluid configuration.
Instead, its leading radial pressure satisfies
$p_r^{\rm eff}\simeq\rho_{\rm eff}$, whereas the two tangential
pressures satisfy
$p_\perp^{\rm eff}\simeq-\rho_{\rm eff}$.
This directional structure is characteristic of the stress tensor associated with a static
radial scalar gradient. 
The anisotropy therefore has a direct origin in the radial
curvature-induced axion profile.

The limiting value in Eq.~\eqref{wa3} should accordingly be interpreted only as the ratio
of the mean principal pressure,
\begin{equation}
\bar p_{\rm eff}
=
\frac{p_r^{\rm eff}+2p_\perp^{\rm eff}}{3},
\end{equation}
to the effective energy density, rather than as the equation-of-state
parameter of an isotropic perfect fluid.
The individual limits~\eqref{wa1} and~\eqref{wa2}, together with the nonvanishing anisotropy~\eqref{wa4},
retain directional information that is lost upon taking the mean-pressure average.

\section{Black hole thermodynamics} % of black holes with axion-cubic-curvature corrections}
\label{sec:bh_chemistry}

We will now evaluate the thermodynamic quantities for the black hole solutions
found in the previous section using the Iyer--Wald formalism~\cite{Iyer:1994ys},
including the perturbative displacement of the event horizon and the corresponding higher-order
corrections to the metric functions.

Evaluating the axion profile at the horizon (\ref{rh1}) yields 
\begin{equation}
\mathfrak{a}(r_H)
=
\frac{1}{72\sqrt{\pi G}}
+
\frac{49\,\alpha_6}{960G^3M^4}
+
\frac{5\sqrt{\pi}\,\alpha_6^2}
{3456\,G^{11/2}M^8}
\ ,
\end{equation}
indicating a curvature-induced near-horizon deformation that remains highly suppressed
for large black holes.
The modified horizon area, from Eq.~\eqref{rh1}, reads
\begin{equation}\label{area-nonzero}
A_H
=
4\pi r_H^2
=
16\pi G^2M^2
\left(
1
-\frac{5\sqrt{\pi}\,\alpha_6}{216G^{5/2}M^4}
-\frac{6933629\pi\,\alpha_6^2}
{71850240G^5M^8}
\right)
\ ,
\end{equation}
which is also modified by terms that are highly suppressed for large $M$. 

The black hole entropy is obtained from the Iyer--Wald Noether-charge formalism~\cite{Iyer:1994ys},
which generalises the Bekenstein--Hawking area law to higher-curvature theories, namely the
surface integral
\begin{equation}
S_{\text{Wald}} = -2\pi \int_{\mathcal{H}}
\frac{\partial \mathcal{L}}{\partial R_{\mu\nu\rho\sigma}}
\,\varepsilon_{\mu\nu}\varepsilon_{\rho\sigma}\, d\Sigma
\ ,
\end{equation}
with $\mathcal H$ the event-horizon cross section, and $\varepsilon_{\mu\nu}$ denoting
its antisymmetric binormal.
For the action (\ref{act1}), one has 
\begin{eqnarray}\label{waldf1}
\frac{\partial\mathcal L}{\partial R_{\mu\nu\rho\sigma}}\equiv {\cal P}^{\mu\nu\rho\sigma}=   \frac{1}{32\pi G}
\left(
g^{\mu\rho}g^{\nu\sigma}
-
g^{\mu\sigma}g^{\nu\rho}
\right)
+
\frac{\alpha_6 G}{2}\,
\mathfrak{a}\,
R^{\mu\nu}{}_{\alpha\beta}
R^{\alpha\beta\rho\sigma}
\ .
\end{eqnarray}
Substituting this expression into the Wald functional and evaluating
it on the corrected horizon yields 
\begin{eqnarray}\label{w1}
S_{\text{Wald}} &=&
4\pi GM^2
\left(
1
+\frac{\sqrt{\pi}\,\alpha_6}
{216\,G^{5/2}M^4}
+\frac{447283\pi\,\alpha_6^2}
{71850240\,G^5M^8}
\right)
\ .
\end{eqnarray}

The Hawking temperature follows consistently from the corrected surface gravity and can be written as 
\begin{equation}\label{hawk1}
T_H =
\frac{1}{8\pi GM}
\left(
1
+\frac{\sqrt{\pi}\,\alpha_6}
{216\,G^{5/2}M^4}
+\frac{1343389\pi\,\alpha_6^2}
{71850240\,G^5M^8}
\right)
\ .
\end{equation}
The heat capacity reads 
\begin{eqnarray}\label{heat-nonzero}
C
&=&
\left(\frac{\partial T_H}{\partial M}\right)^{-1}\nonumber\\&=&-8\pi GM^2
\left(
1
-\frac{5\sqrt{\pi}\,\alpha_6}
{216G^{5/2}M^4}
-\frac{4328749\pi\,\alpha_6^2}
{25660800G^5M^8}
\right)
\ .
\end{eqnarray}
The Schwarzschild value $C=-8\pi GM^2$ is properly recovered for $\alpha_6\to 0$.
Within the perturbative regime continuously connected to the Schwarzschild solution,
the heat capacity therefore remains negative.

The Helmholtz free energy, using Eqs.~\eqref{w1} and~\eqref{hawk1}, reads
\begin{eqnarray}\label{free-nonzero}
F&=&M-T_HS_{\rm Wald}
\nonumber
\\
&=&
\frac{M}{2}
-\frac{\sqrt{\pi}\,\alpha_6}{216\,G^{5/2}M^3}
-\frac{149351\pi\,\alpha_6^2}
{11975040\,G^5M^7}
\ ,
\end{eqnarray}
which also recovers the standard Schwarzschild limit when $\alpha_6\to0$.

We now consider the evaporation process for our black hole solution.
The Hawking luminosity is formally given by
\begin{equation}\label{gb-nonzero}
\frac{dM}{dt}
=
-\sum_{\ell,s}
\int_0^\infty\frac{d\omega}{2\pi}
\frac{\Gamma_{\ell s}(\omega)\,\omega}
{e^{\omega/T_H}\mp1}
\ ,
\end{equation}
where $\Gamma_{\ell s}(\omega)$ denote the greybody factors of the
corresponding perturbation modes in the modified geometry.
A complete calculation of the EFT corrections for these greybody factors 
would require solving wave equations in the deformed background.
We will instead rely on the blackbody approximation, in which
we can write
\begin{equation}\label{bb11}
\frac{dM}{dt}
\simeq
-\sigma A_HT_H^4
\ ,
\end{equation}
where $\sigma$ is the Stefan–Boltzmann constant. 
Combining Eqs.~(\ref{area-nonzero}) and (\ref{hawk1}) gives
\begin{equation}\label{dm-nonzero}
\frac{dM}{dt}
=
-\frac{\sigma}{256\pi^3G^2M^2}
\left(
1
-\frac{\sqrt{\pi}\,\alpha_6}{216G^{5/2}M^4}
-\frac{58579\pi\,\alpha_6^2}
{2661120G^5M^8}
\right)
\ .
\end{equation}
We can next estimate the evaporation time by integrating this expression.

Eq.~\eqref{dm-nonzero} holds for values of the mass for which the EFT expansion
remains perturbatively controlled.
It is therefore appropriate to consider a finite value $M_f>0$ as the final mass,
rather than formally extrapolating the modified solution to $M=0$, that is, we define
\begin{equation}\label{tevap-nonzero}
t_{\rm evap}
\simeq
\int_{M_i}^{M_f}
\frac{dM}{dM/dt}
\ ,
\end{equation}
where $M_i>M_f$ denotes the black hole mass at the beginning of the evaporation.
Since $dM/dt<0$, Eq.~\eqref{tevap-nonzero} gives a positive evaporation time.

Using Eq.~\eqref{dm-nonzero} in the integral~\eqref{tevap-nonzero} in the limit
$\alpha_6\rightarrow0$ yields the usual Schwarzschild behaviour
\begin{equation}\label{tschwarz}
t_{\rm Schw}
=
\frac{256\pi^3G^2}{3\sigma}
\left(
M_i^3-M_f^3
\right)=5120\pi G^2M_i^3
\ ,
\end{equation}
where the last equality is for two bosonic polarizations corresponding to massless gravitons
with $\sigma=\pi^2/60$.

Defining for convenience 
\begin{subequations}\label{PQdef}
\begin{eqnarray}
P_\infty
&=&
-\frac{\sqrt{\pi}}{216\,G^{5/2}}
\ ,
\\
Q_\infty
&=&
-\frac{58579\pi}{2661120G^5}
\ ,
\end{eqnarray}
\end{subequations}
Eq.~(\ref{dm-nonzero}) can be rewritten as
\begin{equation}\label{dmcompact}
\frac{dM}{dt}
=
-\frac{\sigma}{256\pi^3G^2M^2}
\left(
1
+
P_\infty\frac{\alpha_6}{M^4}
+
Q_\infty\frac{\alpha_6^2}{M^8}
\right)
\ .
\end{equation}
On expanding Eq.~\eqref{tevap-nonzero} to second order in $\alpha_6$ therefore yields
\begin{eqnarray}\label{tevap-expanded}
t_{\rm evap}
\simeq
\frac{256\pi^3G^2}{\sigma}
\left[
\frac{M_i^3-M_f^3}{3}
\!+\!
P_\infty\alpha_6
\left(
\frac{1}{M_i}-\frac{1}{M_f}
\right)
\!+\!
\frac{P_\infty^2-Q_\infty}{5}
\alpha_6^2
\left(
\frac{1}{M_f^5}-\frac{1}{M_i^5}
\right)
\right]
\ .
\end{eqnarray}
The inverse powers of $M_f$ in Eq.~(\ref{tevap-expanded}) explicitly show that 
perturbative corrections grow as the black hole mass decreases.
Consequently, the evaporation time \eqref{tevap-expanded} cannot be naively extrapolated to values of $M_f$
for which the EFT corrections become comparable to the leading Schwarzschild
contribution.

It is still interesting to investigate whether higher-curvature corrections can qualitatively
modify the late-time evaporation.
The Hawking temperature~\eqref{hawk1} can be written as
\begin{equation}\label{THcompact}
T_H(M)
=
\frac{1}{8\pi GM}
\left(
1
-
P_\infty\frac{\alpha_6}{M^4}
+
B_\infty\frac{\alpha_6^2}{M^8}
\right)
\ ,
\end{equation}
where
\begin{equation}\label{Binfdef}
B_\infty
\equiv
\frac{1343389\pi}{71850240\,G^5}
\ .
\end{equation}
Introducing $X\equiv\alpha_6/M^4$, a formal solution of $T_H(X_*)=0$ satisfies
\begin{equation}\label{Xstar}
1-P_\infty X_*+B_\infty X_*^2=0.
\end{equation}
For $B_\infty\neq0$, a real positive mass can be associated with such a root only if
$P_\infty^2-4B_\infty\geq0$ and at least one real solution satisfies $\alpha_6/X_*>0$.
The corresponding mass can be written as 
\begin{equation}\label{Mstar}
M_*
=
\left(
\frac{\alpha_6}{X_*}
\right)^{1/4}
\ .
\end{equation}
If $B_\infty=0$, Eq.~\eqref{Xstar} reduces to the linear relation $X_*=1/P_\infty$,
provided $P_\infty\neq0$.
Thus, the modified Hawking temperature admits a zero for finite mass in an appropriate
region of the EFT parameter space, with its existence and location controlled by both
the asymptotic axion background and the EFT coupling.

The associated evaporation dynamics can be characterised directly.
For $P_\infty^2-4B_\infty>0$, the relevant root is simple, $T_H'(M_*)\neq0$,
and the temperature vanishes linearly,
\begin{equation}
T_H(M)
=
T_H'(M_*)(M-M_*)
+
\mathcal O\!\left((M-M_*)^2\right)
\ .
\end{equation}
Provided that the positive horizon area remains finite and analytic at $M_*$,
the blackbody relation~\eqref{bb11} becomes
\begin{equation}\label{criticalflux}
\frac{dM}{dt}
=
-\sigma A_H(M_*)
\left[T_H'(M_*)\right]^4
(M-M_*)^4
+
\mathcal O\!\left((M-M_*)^5\right)
\ .
\end{equation}
A simple zero of the Hawking temperature therefore produces a quartic suppression
of the effective Hawking flux.
Within this approximation, the zero $M_*$ is a candidate asymptotic endpoint
approaching which the mass-loss rate vanishes rapidly.

The status of this candidate endpoint is determined by the perturbative hierarchy.
A useful measure to control the validity of the above results is given by
\begin{equation}\label{epsilonEFT}
\varepsilon_{\rm EFT}(M)
\equiv
\max\left\{
\left|
P_\infty\frac{\alpha_6}{M^4}
\right|,
\left|
B_\infty\frac{\alpha_6^2}{M^8}
\right|
\right\}
\ ,
\end{equation}
since each correction in Eq.~\eqref{THcompact} must remain small with respect to the leading
Schwarzschild contribution.
A parametrically controlled perturbation theory requiring $\varepsilon_{\rm EFT}\ll1$
and cancellation between individually large terms does not restore perturbative control.
At a formal zero $M_*$, Eq.~\eqref{Xstar} and the triangle inequality imply
\begin{equation}\label{EFTboundMstar}
1
\leq
\left|
P_\infty\frac{\alpha_6}{M_*^4}
\right|
+
\left|
B_\infty\frac{\alpha_6^2}{M_*^8}
\right|
\leq
2\,\varepsilon_{\rm EFT}(M_*)
\ .
\end{equation}
Therefore, $\varepsilon_{\rm EFT}(M_*)\geq1/2$ and the modified Hawking temperature~\eqref{THcompact}
cannot vanish while both corrections at $\mathcal O(\alpha_6^2)$ are small compared with
the leading Schwarzschild term.

The bound in Eq.~\eqref{EFTboundMstar} provides a precise interpretation of the remnant-like behaviour.
The $\mathcal O(\alpha_6^2)$ solution exhibits two correlated features:
a zero of the truncated Hawking temperature at finite mass and, for a simple root, a quartic suppression
of the effective Hawking flux.
Within the finite-order evaporation dynamics, these features define $M_*$ as a finite-mass
remnant-like endpoint, characterised by $T_H(M_*)=0$ and an asymptotically vanishing mass-loss rate.
The result therefore represents a qualitative deviation from Schwarzschild evaporation and provides a
concrete mechanism for remnant formation within the perturbative solution through $\mathcal O(\alpha_6^2)$.

The condition $\varepsilon_{\rm EFT}(M_*)\geq1/2$ does not constitute a sharp breakdown criterion
for the underlying EFT and, in particular, does not invalidate the remnant-like endpoint obtained at
$\mathcal O(\alpha_6^2)$.
Rather, it shows that $M_*$ lies in a regime where the higher-curvature sector competes significantly with the
Schwarzschild contribution.
The robustness and quantitative location of this endpoint beyond the present order consequently become
sensitive to terms beyond $\mathcal O(\alpha_6^2)$ in the present perturbative expansion,
as well as to higher-order curvature operators in the EFT.

Higher-order corrections may shift or remove the zero of the Hawking temperature,
but they may also preserve or even enhance the associated flux suppression.~\footnote{We notice that 
the flux~\eqref{criticalflux} is even less suppressed than predicted by the microcanonical
description of the evaporation~\cite{Casadio:1997yv}.} 
In the latter case, a resummation of the derivative expansion or a UV completion
could extend the finite-order remnant-like endpoint into a genuine finite-mass, low- or
zero-temperature remnant.
Establishing its dynamical and thermodynamic stability requires extending
the analysis beyond $\mathcal O(\alpha_6^2)$.

\section{Concluding remarks and outlook}
\label{secfim}

We have investigated black hole solutions and their thermodynamic properties in a gravitational
EFT containing an axion-like field linearly coupled to a cubic Riemann invariant.
This interaction is particularly natural from the EFT perspective, since the same
nontrivial cubic-curvature structure appears in the two-loop counterterm of pure Einstein gravity.
Promoting its coupling to a dynamical axion field provides a direct mechanism through which
spacetime curvature sources a low-energy degree of freedom.

On the Schwarzschild background, the cubic invariant generates the inhomogeneous axion 
given in Eq.~\eqref{ainf}.
The cubic interaction modifies the geometry already at
$\mathcal O(\alpha_6)$ through the nonvanishing asymptotic axion
background, whose value is fixed by horizon regularity according to Eq. (\ref{ainf-horizon}) for the branch considered here.
The curvature-induced radial axion profile contributes to the
gravitational backreaction at $\mathcal O(\alpha_6^2)$, together with
its canonical stress tensor and the corresponding second-order
higher-curvature contributions.
At $\mathcal O(\alpha_6^2)$, the geometry develops comparatively
long-range corrections without generating an additional Newtonian
$1/r$ term, so that the ADM mass  remains unchanged.
Remarkably, for black holes, demanding horizon regularity uniquely determines the resulting
axion profile, including its asymptotic scalar value $\mathfrak a_\infty$ in Eq.~\eqref{ainf-horizon}.
The associated scalar charge is therefore secondary, since it is completely fixed by the
curvature source.
The curvature-induced axion profile can also be viewed as an anisotropic effective source,
with asymptotically opposite radial and tangential pressures. 

Using the Iyer--Wald formalism~\cite{Iyer:1994ys}, we derived the corrections to the horizon geometry,
Wald entropy, Hawking temperature, heat capacity, and Helmholtz free energy.
All quantities smoothly recover their Schwarzschild limits if the EFT coupling is removed.
Within the perturbatively controlled regime connected to the Schwarzschild solution,
the heat capacity remains negative.
Analysing the evaporation further shows that the relative importance of the higher-curvature
corrections grows as the black hole mass decreases.
In the blackbody approximation, these corrections modify both the Hawking luminosity
and the evaporation time.
The late stages of evaporation therefore naturally probe the regime in which terms beyond
those retained in the present perturbative expansion may become relevant, along with
higher-dimensional operators in the gravitational EFT.

An interesting consequence is the possible emergence of remnants.
For suitable EFT parameters, the truncated Hawking temperature admits a formal zero
at finite mass towards which the effective Hawking flux is suppressed according to
Eq.~\eqref{criticalflux}.
Within the finite-order evaporation dynamics, these correlated features define a finite-mass
remnant-like endpoint and provide a concrete mechanism by which the
axion--cubic-curvature interaction can qualitatively alter the late-time evolution
relative to Schwarzschild evaporation.

The formal zero temperature occurs outside the regime controlled by the parameter
$\varepsilon_{\rm EFT}\ll 1$ in Eq.~\eqref{epsilonEFT},
but this does not constitute a sharp breakdown criterion for the underlying EFT
or invalidate the remnant-like endpoint found at the present order.
Rather, it identifies the regime in which the higher-curvature sector competes
significantly with the Schwarzschild contribution and the continuation of the evaporation
dynamics becomes sensitive to terms beyond $\mathcal O(\alpha_6^2)$.
Such corrections may shift or remove that zero, but they may also preserve or enhance
its associated flux suppression.
In the latter case, a resummation of the derivative expansion or a UV completion
could extend the finite-order remnant-like behaviour into a genuine finite-mass,
low- or zero-temperature endpoint.
Determining whether this structure remains valid beyond the present order, and establishing
its dynamical and thermodynamic stability, provides a concrete direction for extending
the present analysis.

Within the regime of validity of the derivative expansion, no additional low-energy
gravitational degrees of freedom are introduced, as discussed in details in Appendix~\ref{sec2}.
The quadratic fluctuation operator remains continuously connected to its Einstein--Hilbert
counterpart and retains the massless graviton pole.
Additional higher-derivative roots occur only at scales beyond the controlled EFT 
regime and should therefore not be interpreted as additional low-energy propagating states.

Moreover, as shown in Appendix~\ref{secPPN}, the cubic-curvature interaction leaves
the standard 1PN parameters unchanged, so that the usual PPN bounds do not directly
constrain $\alpha_6$ at this order.
The first long-range deviation from GR arises instead at 2PN order through the $\mathcal O(\alpha_6^2)$ correction to the spatial metric, for which current VLBI precision yields a projected sensitivity $|\alpha_6|\lesssim 8.3\times10^{130}\ {\rm GeV}^{-1}$ for a Solar-mass source. This sensitivity is parametrically stronger than the bounds inferred from conventional Solar-System observables associated with the more rapidly decaying corrections. Strong-field observables probe a complementary regime, since the black hole geometry is already modified at $\mathcal O(\alpha_6)$ and can therefore directly access the leading cubic-curvature deformation. 
Although the corresponding values of the coupling $\alpha_6$ may appear numerically large, their magnitude alone does not determine the strength of the EFT corrections. Remarkably, a percent-level modification of the photon-sphere radius corresponds to $6.6 \times 10^{135}\, {\rm GeV}^{-1} \lesssim|\alpha_6|\lesssim 5.4\times 10^{141}\ {\rm GeV}^{-1}$ across the stellar-mass interval $5M_\odot\lesssim M\lesssim150M_\odot$, precisely the mass range relevant to compact binaries accessible to current LIGO-Virgo-KAGRA observations. The apparent hierarchy in the dimensionful coupling is compensated by the strong mass suppression entering the dimensionless EFT expansion parameter. Consequently, numerically large values of $\alpha_6$ can still correspond to perturbatively controlled corrections while producing potentially observable strong-field signatures. This makes stellar-mass black holes a particularly interesting arena in which infrared observations can probe the imprint of the cubic-curvature EFT and, indirectly, its underlying UV gravitational dynamics.

\subsection*{{Acknowledgments}} 
R.C.~is partially supported by the INFN grant FLAG, and his work has also been carried out
in the framework of activities of the National Group of Mathematical Physics (GNFM, INdAM).
R.d.R.~thanks the São Paulo Research Foundation -- FAPESP
(Grants No.~2021/01089-1, No.~2025/23004-9, and No.~2026/14943-4)
and the National Council for Scientific and Technological Development -- CNPq
(Grants No.~303742/2023-2 and No.~401567/2023-0), for partial financial support.

\appendix

\section{Perturbative spectrum and higher-derivative modes}
\label{sec2}

Higher-curvature interactions generically raise concerns regarding the emergence of
additional propagating degrees of freedom and the associated Ostrogradsky instabilities
that typically arise in higher-derivative theories of gravity~\cite{Shapiro:2008sf,Shapiro:2015uxa}. 
Hereon, a covariant background-field expansion will be performed around a generic solution
$(\bar g_{\mu\nu},\bar{\mathfrak a})$.
The cubic-curvature operator in Eq.~\eqref{act1} is then treated as a perturbative
higher-derivative correction, with coupling $\alpha_6$.
Its relevance on a given background depends on the combination of this coupling
with the background axion and curvature scales.
The EFT expansion is controlled by requiring higher-curvature contribution remain
perturbatively small relative to the leading two-derivative sector and that the
characteristic curvature and momentum scales remain below the cutoff of the
derivative expansion~\cite{DeFelice:2010hg}.

The higher-curvature operator in Eq.~\eqref{act1} generates higher-derivative terms
when the truncated action is varied exactly.
Since Eq.~\eqref{act1} is an effective action, these equations must be treated perturbatively
in $\alpha_6$ and can be order reduced using the lower-order equations of
motion~\cite{Simon:1990ic}.
It is therefore useful to distinguish the principal symbol of the unreduced higher-derivative
equations from that of the order-reduced EFT equations describing the low-energy spectrum.

Denoting the fluctuation fields collectively by $\Phi^A=\{h_{\mu\nu},\delta\mathfrak a\}$,
the principal symbol of the order-reduced equations is defined by~\cite{Papallo:2017qvl}
\begin{equation}\label{ps1}
P_{\rm red}^{AB}(x,k)
=
\frac{\partial {\cal E}_{\rm red}^{A}}
{\partial(\bar\nabla_\mu\bar\nabla_\nu\Phi^B)}
\,k_\mu k_\nu
\ ,
\end{equation}
where ${\cal E}_{\rm red}^{A}=0$ denotes the perturbatively order-reduced equations of motion.
The derivative in Eq.~\eqref{ps1} is understood in the jet-bundle sense, treating
$\bar\nabla_\mu\bar\nabla_\nu\Phi^B$ as independent symmetric variables~\cite{Kupershmidt:1979rc}.
The principal symbol~\eqref{ps1} determines the characteristic surfaces of the order-reduced
field equations and therefore constrains their hyperbolicity and low-energy propagation.
This analysis should not be confused with a propagator-pole or residue analysis.
In particular, the second-order characteristic structure of the order-reduced system
should not, by itself, be interpreted as establishing that the underlying higher-derivative
theory is ghost free.
Rather, within the regime of validity of the truncated EFT and to the perturbative order
considered here, order reduction ensures that no additional independent higher-derivative
initial data are introduced beyond those associated with the low-energy sector.

The principal symbol is a field-space operator, with indices $A$ and $B$
running over all dynamical fluctuations, and has the block structure (we denote by the
subindex ``red'' the principal symbol of the perturbatively order-reduced equations of motion)
\begin{equation}
P_{\rm red}^{AB}
=
\begin{pmatrix}
P_{\rm red}^{hh} & P_{\rm red}^{h\mathfrak a}
\\
P_{\rm red}^{\mathfrak a h} &
P_{\rm red}^{\mathfrak a\mathfrak a}
\end{pmatrix}
\ .
\end{equation}
Here $P_{\rm red}^{hh}$ governs metric fluctuations, $P_{\rm red}^{\mathfrak a\mathfrak a}$
governs axion fluctuations, and the off-diagonal blocks describe metric--axion mixing.
To investigate whether the higher-curvature interaction introduces an additional low-energy
spin-2 degree of freedom, we focus on the metric sector while keeping in mind that the complete
characteristic problem is defined by the full block operator.
As usual in a diffeomorphism-invariant theory, the metric principal symbol is gauge degenerate.
Statements concerning the physical characteristic structure are therefore understood
after an appropriate gauge fixing or, equivalently, upon restriction to the physical polarization
subspace.

To exhibit explicitly the higher-derivative structure generated by the interaction in Eq.~\eqref{act1},
we take into account the tensor ${\cal P}^{\mu\nu\rho\sigma}$ in Eq.~\eqref{waldf1}. 
For a Lagrangian depending algebraically on the Riemann tensor, the
metric equations contain the contribution
\begin{equation}\label{metricP}
{\cal E}_{\mu\nu}
\supset
-2\nabla_\alpha\nabla_\beta
{\cal P}_{\mu}{}^{\alpha\beta}{}_{\nu}
\ .
\end{equation}
Consequently, the axion-cubic curvature interaction contributes as
\begin{equation}\label{dp1}
{\cal E}_{\mu\nu}^{(6)}
\supset
-{\alpha_6G}\,
\nabla_\alpha\nabla_\beta
\left(
\mathfrak a\,
R_{\mu}{}^{\alpha}{}_{\lambda\kappa}
R^{\lambda\kappa\beta}{}_{\nu}
\right)
\ .
\end{equation}

We now linearly expand around the generic background $(\bar g_{\mu\nu},\bar{\mathfrak a})$
according to
\begin{equation}
g_{\mu\nu}=\bar g_{\mu\nu}+h_{\mu\nu}
\ ,
\qquad
\mathfrak a=\bar{\mathfrak a}+\delta\mathfrak a
\ .
\end{equation}
The terms with the highest number of derivatives of the fluctuations $h_{\mu\nu}$
and $\delta\mathfrak a$ are given by 
\begin{eqnarray}\label{dp2}
\delta{\cal E}_{\mu\nu}^{(6)}
\supset
-{\alpha_6G}\,
\bar\nabla_\alpha\bar\nabla_\beta
\left[
\bar{\mathfrak a}\,
\delta R_{\mu}{}^{\alpha}{}_{\lambda\kappa}
\bar R^{\lambda\kappa\beta}{}_{\nu}
+
\bar{\mathfrak a}\,
\bar R_{\mu}{}^{\alpha}{}_{\lambda\kappa}
\delta R^{\lambda\kappa\beta}{}_{\nu}
+
\delta\mathfrak a\,
\bar R_{\mu}{}^{\alpha}{}_{\lambda\kappa}
\bar R^{\lambda\kappa\beta}{}_{\nu}
\right]
\ ,
\end{eqnarray}
up to terms with fewer derivatives on the fluctuations. 
The two-derivative part of the linearised Riemann tensor reads 
\begin{equation}\label{deltaR}
\left.\delta R_{\mu\nu\rho\sigma}\right|_{\bar\nabla^2h}
=
\frac{1}{2}
\left(
\bar\nabla_\rho\bar\nabla_\nu h_{\mu\sigma}
+
\bar\nabla_\sigma\bar\nabla_\mu h_{\nu\rho}
-
\bar\nabla_\sigma\bar\nabla_\nu h_{\mu\rho}
-
\bar\nabla_\rho\bar\nabla_\mu h_{\nu\sigma}
\right)
\ .
\end{equation}
Eqs.~\eqref{dp2} and~\eqref{deltaR} determine the fourth-order metric principal part
of the unreduced equations.
In the geometric-optics limit, the two outer derivatives in Eq.~\eqref{dp2},
together with the two derivatives contained in $\delta R_{\mu\nu\rho\sigma}$, generate four powers of the
perturbation momentum.
Terms in which derivatives act on $\bar{\mathfrak a}$ or on the background curvature contain fewer
derivatives of $h_{\mu\nu}$ and therefore do not belong to the fourth-order metric principal symbol.

To characterise the spin content of the low-energy metric fluctuations, it is useful to introduce the
Barnes--Rivers projectors in a locally freely falling frame.
At a given spacetime point, Riemann normal coordinates identify the leading Einstein--Hilbert
kinetic operator with its flat-space form.
Defining, as usual, 
\begin{equation}
\theta_{\mu\nu}
=
\bar g_{\mu\nu} - \frac{k_\mu k_\nu}{k^2}
\equiv
\bar g_{\mu\nu} -
\omega_{\mu\nu} 
\ ,
\end{equation}
the identity operator on symmetric tensors decomposes as
\begin{equation}
\mathbf{1} = P^{(2)} + P^{(1)} + P^{(0-s)} + P^{(0-w)}
\ ,
\end{equation}
where the Barnes--Rivers projectors are defined by
\begin{subequations}
\begin{align}
P^{(2)}_{\mu\nu\rho\sigma}
&=
\frac{1}{2}
\left(
\theta_{\mu\rho}\theta_{\nu\sigma}
+
\theta_{\mu\sigma}\theta_{\nu\rho}
\right)
-\frac{1}{3}\theta_{\mu\nu}\theta_{\rho\sigma}
\ ,
\\
P^{(1)}_{\mu\nu\rho\sigma}
&=
\frac{1}{2}
\left(
\theta_{\mu\rho}\omega_{\nu\sigma}
+
\theta_{\mu\sigma}\omega_{\nu\rho}
+
\theta_{\nu\rho}\omega_{\mu\sigma}
+
\theta_{\nu\sigma}\omega_{\mu\rho}
\right)
\ ,
\\
P^{(0-s)}_{\mu\nu\rho\sigma}
&=
\frac{1}{3}\,\theta_{\mu\nu}\theta_{\rho\sigma}
\ ,
\\
P^{(0-w)}_{\mu\nu\rho\sigma}
&=
\omega_{\mu\nu}\omega_{\rho\sigma}
\ .
\end{align}
\end{subequations}
Here $P^{(0-s)}$ and $P^{(0-w)}$ denote the transverse and longitudinal scalar sectors, respectively.
At leading order in the local inertial expansion, and up to the usual gauge-dependent sectors,
the Einstein--Hilbert kinetic operator takes the form
\begin{equation}\label{PEH}
P^{\rm EH}_{\mu\nu\rho\sigma}
=
k^2
\left(
P^{(2)}_{\mu\nu\rho\sigma}
-
P^{(0-s)}_{\mu\nu\rho\sigma}
\right)
\ .
\end{equation}
The Barnes--Rivers basis therefore identifies the transverse-traceless spin-2 sector of the leading
two-derivative theory.
Background curvature and axion gradients can generate additional tensor structures in the
higher-curvature corrections, so the complete curved-background operator need not be exactly
diagonal in the flat-space Barnes--Rivers basis.
Accordingly, the Barnes--Rivers decomposition is used here only as a local probe of the spin
content continuously connected to the flat-space Einstein sector, rather than as an exact
diagonalisation of the complete fluctuation operator on a generic curved background. 
The unreduced equations contain the fourth-order metric principal part displayed explicitly
in Eqs.~\eqref{dp2} and~\eqref{deltaR}.
If the finite higher-derivative truncation were treated as an exact theory, its characteristic
polynomial could have additional roots besides the branch continuously connected to the
Einstein graviton.
Such roots become relevant when the higher-derivative contributions are comparable
to the Einstein--Hilbert kinetic term.
At that scale, however, the derivative expansion underlying Eq.~\eqref{act1} is no longer
parametrically controlled, and higher-order operators omitted from the truncated action can
contribute at comparable order.

The EFT equations employed here are instead treated perturbatively and order reduced.
At each order in $\alpha_6$, higher derivatives generated by Eq.~\eqref{dp1} are eliminated
perturbatively using the lower-order Einstein-axion equations.
The resulting low-energy equations remain second order in the perturbative EFT description.
Consequently, the higher-derivative terms do not require additional independent spin-2
initial data associated with the formal higher-derivative branches of the unreduced equations.
The physical spin-2 sector is continuously connected to the Einstein graviton sector,
although the background-dependent EFT corrections may modify its characteristic propagation.

This conclusion should be understood strictly within the domain of validity of the EFT.
Formal additional high-frequency solutions obtained by solving the finite unreduced higher-derivative
equations exactly are not predictions of the low-energy theory when their characteristic momentum
approaches the cutoff, because operators omitted from Eq.~\eqref{act1} can no longer be
neglected~\cite{Kuntz:2019qcf}.
We therefore do not claim that the finite higher-derivative action, regarded as an exact theory
at arbitrary momentum, is nonperturbatively ghost-free.
Rather, after consistent perturbative order reduction, the operator in Eq.~\eqref{act1} introduces no
additional independent low-energy spin-2 degree of freedom beyond the graviton already present
in Einstein gravity.
The axion remains the scalar degree of freedom already present in the two-derivative
Einstein-axion theory.

\section{Effective anisotropic fluid}
\label{appb1}

To characterise the effective matter content corresponding to the higher-curvature deformation
of the geometry, it is useful to recast the modified Einstein equations in terms of an effective
anisotropic source.
For the metric~\eqref{br1} and \eqref{ar1}, we define the effective energy density and principal
pressures directly from the Einstein tensor as
\begin{eqnarray}\label{rhoeff1}
\rho_{\rm eff}
&=&
\frac{\alpha_6}{9\sqrt{\pi G}}
\left(
\frac{45\,G^3M^2}{r^8}
-
\frac{98\,G^4M^3}{r^9}
\right)
\nonumber\\
&&
\!\!\!+\alpha_6^2
\left(
\frac{1}{1152\,G^4M^6r^4}
+\frac{1}{576\,G^3M^5r^5}
+\frac{1}{288\,G^2M^4r^6}
+\frac{1}{144\,GM^3r^7}
+\frac{1}{72\,M^2r^8}
\right.
\nonumber\\
&&\left.\qquad
+\frac{793\,G}{36Mr^9}
-\frac{307\,G^2}{18r^{10}}
-\frac{103\,G^3M}{9r^{11}}
-\frac{100\,G^4M^2}{9r^{12}}
-\frac{112\,G^5M^3}{9r^{13}}
\right.
\nonumber\\
&&\left.\qquad
+\frac{856\,G^6M^4}{45\,r^{14}}
-\frac{74488\,G^7M^5}{81\,r^{15}}
\right)
\ .
\end{eqnarray}
Similarly, the radial pressure is given by
\begin{eqnarray}\label{preff1}
p_r^{\rm eff}
&=&
\frac{\alpha_6}{9\sqrt{\pi G}}
\left(
\frac{9\,G^3M^2}{r^8}
-
\frac{10\,G^4M^3}{r^9}
\right)
\nonumber\\
&&
+\alpha_6^2
\left(
\frac{1}{1152\,G^4M^6r^4}
+\frac{1}{576\,G^3M^5r^5}
+\frac{1}{288\,G^2M^4r^6}
+\frac{1}{144\,GM^3r^7}
+\frac{1}{72\,M^2r^8}
\right.
\nonumber\\
&&\qquad\left.
+\frac{121\,G}{36Mr^9}
+\frac{5\,G^2}{18r^{10}}
+\frac{17\,G^3M}{9r^{11}}
+\frac{4\,G^4M^2}{r^{12}}
-\frac{736\,G^5M^3}{15\,r^{13}}
\right.
\nonumber\\
&&\qquad\left.
+\frac{2392\,G^6M^4}{9\,r^{14}}
-\frac{23576\,G^7M^5}{81\,r^{15}}
\right)
\ ,
\end{eqnarray}
whereas the tangential pressure becomes
\begin{eqnarray}\label{pteff1}
p_\perp^{\rm eff}
&=&
-\frac{\alpha_6}{9\sqrt{\pi G}}
\left(
\frac{27\,G^3M^2}{r^8}
-
\frac{62\,G^4M^3}{r^9}
\right)
\nonumber\\
&&
-\alpha_6^2
\left(
\frac{1}{1152\,G^4M^6r^4}
+\frac{1}{576\,G^3M^5r^5}
+\frac{1}{288\,G^2M^4r^6}
+\frac{1}{144\,GM^3r^7}
+\frac{1}{72\,M^2r^8}
\right.
\nonumber\\
&&\qquad\left.
+\frac{421\,G}{36Mr^9}
-\frac{211\,G^2}{18r^{10}}
-\frac{79\,G^3M}{9r^{11}}
-\frac{88\,G^4M^2}{9r^{12}}
-\frac{4888\,G^5M^3}{15\,r^{13}}
\right.
\nonumber\\
&&\qquad\left.
+\frac{68104\,G^6M^4}{45\,r^{14}}
-\frac{176680\,G^7M^5}{81\,r^{15}}
\right)
\ .
\end{eqnarray}
These quantities incorporate both the higher-curvature contribution and the
backreaction associated with the curvature-induced axion profile~\eqref{ainf}
with \eqref{ainf-horizon},
and provide a convenient description of the radial structure and anisotropy
generated by the EFT corrections.

\section{Post-Newtonian constraints from Solar-System tests}
\label{secPPN}

The metric functions~\eqref{br1} and~\eqref{ar1} have an asymptotic regime
in which the parametrised post-Newtonian (PPN) expansion~\cite{Will:2014kxa}
can be employed to bound the EFT parameter $\alpha_6$, namely 
\begin{eqnarray}
B(r)
&=&
1-\frac{2GM}{r}
+\mathcal{O}(r^{-3})
\ ,
\label{br2}
\\
A(r)
&=&
1-\frac{2GM}{r}
+\frac{\pi\alpha_6^2}{144G^3M^6r^2}
+\mathcal{O}(r^{-3})
\ ,
\label{ar2}
\end{eqnarray}
where we displayed the long-range contributions relevant for the standard 1PN
parameters and the leading 2PN correction in the spatial metric.
In particular, terms linear in $\alpha_6$ in Eqs.~\eqref{br1} and~\eqref{ar1} do not
affect the 1PN coefficients since they decay as $r^{-6}$ and $r^{-7}$.

To compare with standard analysis, we introduce the isotropic radial coordinate 
\begin{equation}
\frac{d\rho}{\rho}
=
\frac{dr}{r\sqrt{A(r)}}
\ ,
\end{equation}
which yields 
\begin{eqnarray}
r
=
\rho+GM
+
\frac{1}{4\rho}
\left(
G^2M^2-\frac{\pi\alpha_6^2}{144G^3M^6}
\right)
+\mathcal{O}(\rho^{-2})
\ .
\label{isotrans}
\end{eqnarray}
Substituting Eq.~\eqref{isotrans} into the metric function~\eqref{br2} gives  
\begin{eqnarray}
B(\rho)
=
1-\frac{2GM}{\rho}
+\frac{2G^2M^2}{\rho^2}
+\mathcal{O}(\rho^{-3})
\ ,
\end{eqnarray}
and, defining $U={GM}/{\rho}$, one finds 
\begin{eqnarray}
g_{tt}
=
-B(\rho)
=
-1+2U-2U^2+\mathcal{O}(U^3)
\ .
\label{gttPPN}
\end{eqnarray}
Comparison with the standard PPN form~\cite{Will:2014kxa} 
\begin{equation}
g_{tt}=
-1+2U-2\beta_{\rm PPN}U^2+\mathcal{O}(U^3)
\end{equation}
simply yields $\beta_{\rm PPN}=1$ like in GR. 

The spatial components of the metric~\eqref{ans1} in isotropic coordinates read 
\begin{eqnarray}
g_{ij}
=
\frac{r^2(\rho)}{\rho^2}\,\delta_{ij}
=
\left[
1+2U+
\left(
\frac{3}{2}-\zeta_6
\right)U^2
\right]\delta_{ij}
+
\mathcal{O}(U^3)
\ ,
\label{2PNmetric}
\end{eqnarray}
where the parameter
\begin{eqnarray}\label{z6}
\zeta_6
\equiv
\frac{\pi\alpha_6^2}{288G^5M^8}
\end{eqnarray}
characterises the $\mathcal{O}(U^2)$ spatial correction and should not be confused
with the standard PPN parameters conventionally denoted by $\zeta_i$.
Comparison with~\cite{Will:2014kxa}
\begin{equation}
g_{ij}
=
\left(
1+2\gamma_{\rm PPN}U+\cdots
\right)\delta_{ij}
\ ,
\end{equation}
implies that $\gamma_{\rm PPN}=1$ again like in GR.

Since both 1PN parameters $\gamma_{\rm PPN}=\beta_{\rm PPN}=1$,
the usual 1PN Cassini and perihelion observations do not constrain $\alpha_6$.
Instead, the leading long-range corrections quadratic in $\alpha_6$ appear at
2PN order in the spatial metric~\cite{Richter:1982zz} and is parametrised here by
$\zeta_6$. 

We next consider light deflection in the Solar system.
The effective optical refractive index, taking into account Eqs.~\eqref{gttPPN} and \eqref{2PNmetric},
can be written in isotropic coordinates as
\begin{eqnarray}
n(\rho)
=
\sqrt{
\frac{r^2(\rho)}
{\rho^2 B(\rho)}
}=
1+2U+
\frac12\left(
\frac{7}{2}-{\zeta_6}
\right)U^2
+\mathcal{O}(U^3)
\ .
\end{eqnarray}
The contribution generated specifically by the cubic-curvature correction is therefore given by
\begin{eqnarray}
\delta n_6
=
-\frac{\zeta_6\,G^2M^2}{2\rho^2}
\ .
\label{deltaindex}
\end{eqnarray}
For a light ray with impact parameter $b$, we may evaluate this additional 2PN contribution
along the unperturbed trajectory, $\rho^2=b^2+z^2$.
To first order in $\delta n_6$, the corresponding additional deflection reads 
\begin{eqnarray}
\left|\delta\theta_6\right|
=
\left|
\int_{-\infty}^{+\infty}
\frac{\partial\,\delta n_6}{\partial b}\,dz
\right|
=
|\zeta_6|G^2M^2b
\int_{-\infty}^{+\infty}
\frac{dz}{(b^2+z^2)^2}
= 
\frac{\pi}{2}
|\zeta_6|
\left(
\frac{GM}{b}
\right)^2
\ .
\label{zetadeflection}
\end{eqnarray}
For a ray grazing the Sun, $M=M_\odot$ and $b\simeq R_\odot$, and using
${GM_\odot}/{R_\odot} \simeq 2.12\times10^{-6}$ finally yields 
\begin{eqnarray}
\left|\delta\theta_6\right|
\simeq
1.46\,|\zeta_6|\ \mu{\rm as}
\ .
\label{VLBIangle}
\end{eqnarray}

A bound for the EFT can then be obtained by requiring that
the additional deflection~\eqref{VLBIangle} does not exceed the angular uncertainty
$\sigma_\theta$ associated with present VLBI measurements.
Taking the VLBA determination $\gamma_{\rm PPN}=0.9998\pm0.0003$~\cite{Fomalont:2009zg}
as a representative experimental uncertainty $\sigma_\gamma\simeq3\times10^{-4}$ and,
using the GR deflection angle $\theta_{\rm GR}\simeq1.75\,$arcsec~\cite{Will:2014kxa}, the
corresponding angular uncertainty 
$\sigma_\theta\simeq {\sigma_\gamma\,\theta_{\rm GR}}/{2}\simeq2.6\times10^2\,\mu{\rm as}$. 
Requiring $|\delta\theta_6|\lesssim \sigma_\theta$ therefore gives the conservative upper bound
\begin{eqnarray}
|\zeta_6|
\lesssim
1.8\times10^2
\ .
\label{zeta6bound}
\end{eqnarray}
This bound can be expressed directly in terms of the EFT coupling $\alpha_6$. 
For a Solar-mass source, one obtains
\begin{eqnarray}
|\alpha_6|
\lesssim
12\,G^{5/2}M_\odot^4
\sqrt{\frac{2}{\pi}|\zeta_6|}=8.3\times10^{130}\ {\rm GeV}^{-1}
\ .
\label{alpha6PPNbound}
\end{eqnarray} 

\begin{figure}[t]
\centering
\includegraphics[width=10cm]{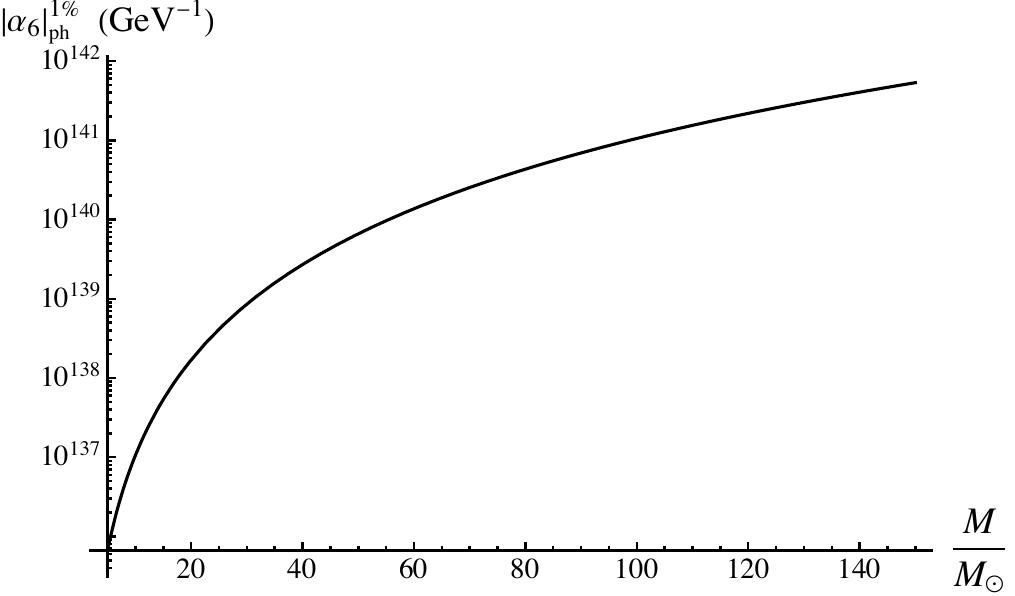}
\captionsetup{
  width=\linewidth,
  justification=raggedright,
  singlelinecheck=false
}
\caption{Upper bound on $|\alpha_6|$ corresponding to $1\%$ modification of the
photon-sphere radius over the stellar-mass range accessible to ground-based gravitational-wave
observations.}
\label{alpha6mass}
\end{figure}

It is useful to further analyse the EFT corrections at 2PN order in the context of both strong-field
observables and conventional Solar-System tests.
For example, in the black hole geometry~\eqref{br1} and~\eqref{ar1}, the radius of the unstable
photon sphere is given by
\begin{eqnarray}
r_{\rm ph}
=
r_{\rm ph}^{\rm GR}
-
\frac{20\sqrt{\pi}\,\alpha_6}
{2187\,G^{3/2}M^3}
+\mathcal O(\alpha_6^2)
\ ,
\end{eqnarray} 
where the Schwarzschild value $r_{\rm ph}^{\rm GR}=3GM$. 
Thus, requiring that the radius of the photon sphere is not changed by more than $1\%$,
that is $\delta_{\rm ph}\equiv |\Delta r_{\rm ph}|/r_{\rm ph}^{\rm GR}\lesssim 0.01$, would correspond to
\begin{eqnarray}
|\alpha_6|_{\rm ph}
\lesssim
\frac{6521}{20\sqrt{\pi}}\,
\delta_{\rm ph}\,
\frac{G^2M^4}{M_{\rm Pl}}
=
1.85\,
\frac{G^2M^4}{M_{\rm Pl}}
\equiv
|\alpha_6|_{\rm ph}^{1\%}
\ .
\label{pha1}
\end{eqnarray} 
Black holes in mergers accessible to current gravitational-wave observations
from LIGO, Virgo, and KAGRA~\cite{KAGRA:2021vkt} lie in the range
$5M_\odot\lesssim M\lesssim150M_\odot$.
For such masses, Eq.~\eqref{pha1} yields upper bounds
\begin{equation}
6.6\times10^{135}\ {\rm GeV}^{-1}
\lesssim
|\alpha_6|_{\rm ph}^{1\%}
\lesssim
5.4\times10^{141}\ {\rm GeV}^{-1}
\ ,
\end{equation} 
which are shown in Fig.~\ref{alpha6mass}.

For comparison, conventional Solar-System observables provide complementary weak-field
probes of $\alpha_6$ through the asymptotic exterior solution.
The perihelion precession of Mercury and the Galileo gravitational-redshift measurement
imply an upper bound $|\alpha_6|\lesssim 1.34\times10^{167}\,{\rm GeV}^{-1}$ and
$3.75\times10^{165}\ {\rm GeV}^{-1}$, respectively.
Shapiro time-delay and solar light deflection improve the corresponding sensitivities to
$1.20\times10^{159}\,{\rm GeV}^{-1}$ and $5.37\times10^{158}\,{\rm GeV}^{-1}$,
while geodetic precession reaches $3.42\times10^{155}\,{\rm GeV}^{-1}$.
These bounds are therefore parametrically weaker than the projected
quadratic 2PN sensitivity in Eq.~\eqref{alpha6PPNbound}. 

Despite their apparently enormous magnitude, these scales comply with the perturbative
character of the EFT defined by the action~\eqref{act1}.
A large numerical value of $\alpha_6$ does not by itself measure the strength of
higher-curvature corrections.
The relevant estimator is the dimensionless EFT expansion parameter
$\varepsilon_{\rm EFT}$ defined in Eq.~\eqref{epsilonEFT}.
Comparing the cubic-curvature interaction with the Einstein term gives 
$\varepsilon_{\rm EFT} = \frac{\pi}{216} |c_6|M_{\rm Pl}^4/{M}^4\ll1$, 
which shows explicitly the strong suppression caused by the gravitational mass $M$.

Such considerations are particularly relevant for astrophysical black holes.
For example, considering  $M=100\,M_\odot$, Eq.~\eqref{pha1}
yields $|\alpha_6|_{\rm ph}^{1\%}\simeq 1.06\times10^{141}\,{\rm GeV}^{-1}$. 
Evaluating the EFT parameter at this characteristic coupling
gives $\varepsilon_{\rm EFT} = 3.4 \times 10^{-3}$.
Therefore, the condition~\eqref{pha1} for
sub-percent photon-sphere modifications is within the range of validity of the EFT.
This identifies an astrophysical regime in which potentially measurable deviations
from the Schwarzschild geometry may coexist with perturbative control of the
cubic-curvature interaction.
These results illustrate how higher-curvature operators in the gravitational EFT 
an encode signatures of UV physics in macroscopic black hole observables.
Black-hole photon spheres may therefore provide an infrared probe of UV
gravitational dynamics.

\bibliography{exotic.bib}

@article{Sotiriou:2015pka,
  author  = "Sotiriou, Thomas P.",
  title   = "{Black Holes and Scalar Fields}",
  journal = "Class. Quant. Grav.",
  volume  = "32",
  number  = "21",
  pages   = "214002",
  year    = "2015",
  doi     = "10.1088/0264-9381/32/21/214002",
  eprint  = "1505.00248",
  archivePrefix = "arXiv",
  primaryClass  = "gr-qc"
}

@article{Sotiriou:2013qea,
  author  = "Sotiriou, Thomas P. and Zhou, Shuang-Yong",
  title   = "{Black hole hair in generalized scalar-tensor gravity}",
  journal = "Phys. Rev. Lett.",
  volume  = "112",
  pages   = "251102",
  year    = "2014",
  doi     = "10.1103/PhysRevLett.112.251102",
  eprint  = "1312.3622",
  archivePrefix = "arXiv",
  primaryClass  = "gr-qc"
}

@article{Bai:2021nrs,
    author = "Bai, Yang and Du, Xiaolong and Hamada, Yuta",
    title = "{Diluted axion star collisions with neutron stars}",
    eprint = "2109.01222",
    archivePrefix = "arXiv",
    primaryClass = "astro-ph.CO",
    doi = "10.1088/1475-7516/2022/01/041",
    journal = "JCAP",
    volume = "01",
    number = "01",
    pages = "041",
    year = "2022"
}

@article{Giddings:1987cg,
    author = "Giddings, Steven B. and Strominger, Andrew",
    title = "{Axion Induced Topology Change in Quantum Gravity and String Theory}",
    reportNumber = "HUTP-87-A067",
    doi = "10.1016/0550-3213(88)90446-4",
    journal = "Nucl. Phys. B",
    volume = "306",
    pages = "890--907",
    year = "1988"
}

@article{Chen:2023vet,
    author = "Chen, Jun-Ru and Huang, Long-Xing and Zhao, Li and Wang, Yong-Qiang",
    title = "{Tidal Love numbers of axion stars}",
    eprint = "2311.11830",
    archivePrefix = "arXiv",
    primaryClass = "gr-qc",
    doi = "10.1103/PhysRevD.109.104078",
    journal = "Phys. Rev. D",
    volume = "109",
    number = "10",
    pages = "104078",
    year = "2024"
}

@article{Gorghetto:2024vnp,
    author = "Gorghetto, Marco and Hardy, Edward and Villadoro, Giovanni",
    title = "{More axion stars from strings}",
    eprint = "2405.19389",
    archivePrefix = "arXiv",
    primaryClass = "hep-ph",
    reportNumber = "DESY-24-075",
    doi = "10.1007/JHEP08(2024)126",
    journal = "JHEP",
    volume = "08",
    pages = "126",
    year = "2024"
}

@article{Shapiro:2008sf,
    author = "Shapiro, Ilya L.",
    title = "{Effective Action of Vacuum: Semiclassical Approach}",
    eprint = "0801.0216",
    archivePrefix = "arXiv",
    primaryClass = "gr-qc",
    doi = "10.1088/0264-9381/25/10/103001",
    journal = "Class. Quant. Grav.",
    volume = "25",
    pages = "103001",
    year = "2008"
}

@article{Shapiro:2015uxa,
    author = "Shapiro, Ilya L.",
    title = "{Counting ghosts in the {\textquotedblleft}ghost-free{\textquotedblright} non-local gravity}",
    eprint = "1502.00106",
    archivePrefix = "arXiv",
    primaryClass = "hep-th",
    doi = "10.1016/j.physletb.2015.03.037",
    journal = "Phys. Lett. B",
    volume = "744",
    pages = "67--73",
    year = "2015"
}

@article{Papallo:2017qvl,
    author = "Papallo, Giuseppe and Reall, Harvey S.",
    title = "{On the local well-posedness of Lovelock and Horndeski theories}",
    eprint = "1705.04370",
    archivePrefix = "arXiv",
    primaryClass = "gr-qc",
    doi = "10.1103/PhysRevD.96.044019",
    journal = "Phys. Rev. D",
    volume = "96",
    number = "4",
    pages = "044019",
    year = "2017"
}

@article{DeFelice:2010hg,
    author = "De Felice, Antonio and Tanaka, Takahiro",
    title = "{Inevitable ghost and the degrees of freedom in f(R,G) gravity}",
    eprint = "1006.4399",
    archivePrefix = "arXiv",
    primaryClass = "astro-ph.CO",
    doi = "10.1143/PTP.124.503",
    journal = "Prog. Theor. Phys.",
    volume = "124",
    pages = "503--515",
    year = "2010"
}

@article{Kupershmidt:1979rc,
    author = "Kupershmidt, B. A.",
    title = "{Geometry of jet bundles and the structure of Lagrangian and Hamiltonian formalisms}",
    journal = "Lect. Notes Math.",
    volume = "775",
    pages = "162--218",
    year = "1980"
}

@article{Kuntz:2019qcf,
    author = "Kuntz, Iber{\^e}",
    title = "{Exorcising ghosts in quantum gravity}",
    eprint = "1909.11072",
    archivePrefix = "arXiv",
    primaryClass = "hep-th",
    doi = "10.1140/epjp/s13360-020-00875-x",
    journal = "Eur. Phys. J. Plus",
    volume = "135",
    number = "10",
    pages = "859",
    year = "2020"
}

@article{Simon:1990ic,
    author = "Simon, Jonathan Z.",
    title = "{Higher Derivative Lagrangians, Nonlocality, Problems and Solutions}",
    reportNumber = "UCSB-TH-89-50",
    doi = "10.1103/PhysRevD.41.3720",
    journal = "Phys. Rev. D",
    volume = "41",
    pages = "3720",
    year = "1990"
}

@article{KAGRA:2021vkt,
    author = "Abbott, R. and others",
    collaboration = "KAGRA, VIRGO, LIGO Scientific",
    title = "{GWTC-3: Compact Binary Coalescences Observed by LIGO and Virgo during the Second Part of the Third Observing Run}",
    eprint = "2111.03606",
    archivePrefix = "arXiv",
    primaryClass = "gr-qc",
    reportNumber = "LIGO-P2000318",
    doi = "10.1103/PhysRevX.13.041039",
    journal = "Phys. Rev. X",
    volume = "13",
    number = "4",
    pages = "041039",
    year = "2023"
}

@article{Buoninfante:2025neu,
    author = "Buoninfante, Luca and Giusti, Andrea and Held, Aaron and Knorr, Benjamin and Platania, Alessia",
    title = "{Higher derivatives in quantum gravity: theory, tests, phenomenology}",
    doi = "10.1140/epjp/s13360-025-06690-6",
    journal = "Eur. Phys. J. Plus",
    volume = "140",
    number = "8",
    pages = "772",
    year = "2025"
}

@article{Richter:1982zz,
    author = "Richter, Gary W. and Matzner, Richard A.",
    title = "{Second-order contributions to gravitational deflection of light in the parametrized post-Newtonian formalism}",
    doi = "10.1103/PhysRevD.26.1219",
    journal = "Phys. Rev. D",
    volume = "26",
    pages = "1219--1224",
    year = "1982"
}

@article{Will:2014kxa,
    author = "Will, Clifford M.",
    title = "{The Confrontation between General Relativity and Experiment}",
    eprint = "1403.7377",
    archivePrefix = "arXiv",
    primaryClass = "gr-qc",
    doi = "10.12942/lrr-2014-4",
    journal = "Living Rev. Rel.",
    volume = "17",
    pages = "4",
    year = "2014"
}

@article{Fomalont:2009zg,
    author = "Fomalont, E. and Kopeikin, S. and Lanyi, G. and Benson, J.",
    title = "{Progress in Measurements of the Gravitational Bending of Radio Waves Using the VLBA}",
    eprint = "0904.3992",
    archivePrefix = "arXiv",
    primaryClass = "astro-ph.CO",
    doi = "10.1088/0004-637X/699/2/1395",
    journal = "Astrophys. J.",
    volume = "699",
    pages = "1395--1402",
    year = "2009"
}

@article{Burgess:2003jk,
    author = "Burgess, C. P.",
    title = "{Quantum gravity in everyday life: General relativity as an effective field theory}",
    eprint = "gr-qc/0311082",
    archivePrefix = "arXiv",
    doi = "10.12942/lrr-2004-5",
    journal = "Living Rev. Rel.",
    volume = "7",
    pages = "5--56",
    year = "2004"
}

@article{Barranco:2012ur,
  author = {Barranco, J. and Monteverde, A. Carrillo and Delepine, D.},
  title = {Can the dark matter halo be a collisionless ensemble of axion stars?},
  eprint = {1212.2254},
  archivePrefix = {arXiv},
  primaryClass = {astro-ph.CO},
  doi = {10.1103/PhysRevD.87.103011},
  journal = {Phys. Rev. D},
  volume = {87},
  pages = {103011},
  year = {2013}
}

@article{Bueno:2023jtc,
    author = "Bueno, Pablo and Cano, Pablo A. and Hennigar, Robie A.",
    title = "{On the stability of Einsteinian cubic gravity black holes in EFT}",
    eprint = "2306.02924",
    archivePrefix = "arXiv",
    primaryClass = "hep-th",
    doi = "10.1088/1361-6382/ad4f41",
    journal = "Class. Quant. Grav.",
    volume = "41",
    number = "13",
    pages = "137001",
    year = "2024"
}

@article{deFreitasPacheco:2018gpi,
    author = "de Freitas Pacheco, J. A. and Carneiro, S. and Fabris, J. C.",
    title = "{Gravitational waves from binary axionic black holes}",
    eprint = "1811.02289",
    archivePrefix = "arXiv",
    primaryClass = "gr-qc",
    doi = "10.1140/epjc/s10052-019-6940-z",
    journal = "Eur. Phys. J. C",
    volume = "79",
    number = "5",
    pages = "426",
    year = "2019"
}

@article{Levkov:2018kau,
  author = {Levkov, D. G. and Panin, A. G. and Tkachev, I. I.},
  title = {Gravitational Bose-Einstein condensation in the kinetic regime},
  eprint = {1804.05857},
  archivePrefix = {arXiv},
  primaryClass = {astro-ph.CO},
  doi = {10.1103/PhysRevLett.121.151301},
  journal = {Phys. Rev. Lett.},
  volume = {121},
  number = {15},
  pages = {151301},
  year = {2018}
}

@article{daRocha:2012pt,
    author = "da Rocha, Roldao and Hoff da Silva, J. M.",
    title = "{Black string corrections in variable tension braneworld scenarios}",
    eprint = "1202.1256",
    archivePrefix = "arXiv",
    primaryClass = "gr-qc",
    doi = "10.1103/PhysRevD.85.046009",
    journal = "Phys. Rev. D",
    volume = "85",
    pages = "046009",
    year = "2012"
}

@article{Hogan:1988mp,
    author = "Hogan, C. J. and Rees, M. J.",
    title = "{AXION MINICLUSTERS}",
    doi = "10.1016/0370-2693(88)91655-3",
    journal = "Phys. Lett. B",
    volume = "205",
    pages = "228--230",
    year = "1988"
}

@article{Casadio:2023mgl,
    author = "Casadio, R. and da Rocha, R.",
    title = "{Axion stars in MGD background}",
    eprint = "2305.15752",
    archivePrefix = "arXiv",
    primaryClass = "hep-th",
    doi = "10.1140/epjc/s10052-023-11731-4",
    journal = "Eur. Phys. J. C",
    volume = "83",
    number = "6",
    pages = "537",
    year = "2023"
}

@article{Abdalla:2009pg,
    author = "Abdalla, M. C. B. and Hoff da Silva, J. M. and da Rocha, R.",
    title = "{Notes on the Two-brane Model with Variable Tension}",
    eprint = "0907.1321",
    archivePrefix = "arXiv",
    primaryClass = "hep-th",
    doi = "10.1103/PhysRevD.80.046003",
    journal = "Phys. Rev. D",
    volume = "80",
    pages = "046003",
    year = "2009"
}

@article{Ballesteros:2023iqb,
    author = "Ballesteros, Romina and G{\'o}mez-Fayr{\'e}n, Carmen and Ort{\'\i}n, Tom{\'a}s and Zatti, Matteo",
    title = "{On scalar charges and black hole thermodynamics}",
    eprint = "2302.11630",
    archivePrefix = "arXiv",
    primaryClass = "hep-th",
    reportNumber = "IFT-UAM/CSIC-23-018",
    doi = "10.1007/JHEP05(2023)158",
    journal = "JHEP",
    volume = "05",
    pages = "158",
    year = "2023"
}

@article{Cicoli:2026fqp,
    author = "Cicoli, Michele and others",
    title = "{Axions at the meV crossroads: theory, cosmology, astrophysics, and experiments}",
    eprint = "2603.18167",
    archivePrefix = "arXiv",
    primaryClass = "hep-ph",
    reportNumber = "LAPTH-006/26, BARI-TH/785-26",
    doi = "10.1088/1475-7516/2026/07/060",
    journal = "JCAP",
    volume = "07",
    pages = "060",
    year = "2026"
}

@article{Calmet:2021stu,
    author = "Calmet, Xavier and Casadio, Roberto and Hsu, Stephen D. H. and Kuipers, Folkert",
    title = "{Quantum Hair from Gravity}",
    eprint = "2110.09386",
    archivePrefix = "arXiv",
    primaryClass = "hep-th",
    doi = "10.1103/PhysRevLett.128.111301",
    journal = "Phys. Rev. Lett.",
    volume = "128",
    number = "11",
    pages = "111301",
    year = "2022"
}

@article{Cavalcanti:2016mbe,
    author = "Cavalcanti, R. T. and da Silva, A. Goncalves and da Rocha, Roldao",
    title = "{Strong deflection limit lensing effects in the minimal geometric deformation and Casadio{\textendash}Fabbri{\textendash}Mazzacurati solutions}",
    eprint = "1605.01271",
    archivePrefix = "arXiv",
    primaryClass = "gr-qc",
    doi = "10.1088/0264-9381/33/21/215007",
    journal = "Class. Quant. Grav.",
    volume = "33",
    number = "21",
    pages = "215007",
    year = "2016"
}

@article{Calmet:2021lny,
    author = "Calmet, Xavier and Kuipers, Folkert",
    title = "{Quantum gravitational corrections to the entropy of a Schwarzschild black hole}",
    eprint = "2108.06824",
    archivePrefix = "arXiv",
    primaryClass = "hep-th",
    doi = "10.1103/PhysRevD.104.066012",
    journal = "Phys. Rev. D",
    volume = "104",
    number = "6",
    pages = "066012",
    year = "2021"
}

@article{Bueno:2016ypa,
    author = "Bueno, Pablo and Cano, Pablo A. and Min, Vincent S. and Visser, Manus R.",
    title = "{Aspects of general higher-order gravities}",
    eprint = "1610.08519",
    archivePrefix = "arXiv",
    primaryClass = "hep-th",
    reportNumber = "IFT-UAM-CSIC-16-108",
    doi = "10.1103/PhysRevD.95.044010",
    journal = "Phys. Rev. D",
    volume = "95",
    number = "4",
    pages = "044010",
    year = "2017"
}

@article{Doneva:2017bvd,
    author = "Doneva, Daniela D. and Yazadjiev, Stoytcho S.",
    title = "{New Gauss-Bonnet Black Holes with Curvature-Induced Scalarization in Extended Scalar-Tensor Theories}",
    eprint = "1711.01187",
    archivePrefix = "arXiv",
    primaryClass = "gr-qc",
    doi = "10.1103/PhysRevLett.120.131103",
    journal = "Phys. Rev. Lett.",
    volume = "120",
    number = "13",
    pages = "131103",
    year = "2018"
}

@article{Bueno:2018xqc,
    author = "Bueno, Pablo and Cano, Pablo A. and Ruip{\'e}rez, Alejandro",
    title = "{Holographic studies of Einsteinian cubic gravity}",
    eprint = "1802.00018",
    archivePrefix = "arXiv",
    primaryClass = "hep-th",
    doi = "10.1007/JHEP03(2018)150",
    journal = "JHEP",
    volume = "03",
    pages = "150",
    year = "2018"
}

@article{Agrawal:2024ejr,
    author = "Agrawal, Prateek and Nee, Michael and Reig, Mario",
    title = "{Axion couplings in heterotic string theory}",
    eprint = "2410.03820",
    archivePrefix = "arXiv",
    primaryClass = "hep-ph",
    doi = "10.1007/JHEP02(2025)188",
    journal = "JHEP",
    volume = "02",
    pages = "188",
    year = "2025"
}

@article{vandeVen:1991gw,
    author = "van de Ven, A. E. M.",
    title = "{Two loop quantum gravity}",
    reportNumber = "DESY-91-115, ITP-SB-91-52",
    doi = "10.1016/0550-3213(92)90011-Y",
    journal = "Nucl. Phys. B",
    volume = "378",
    pages = "309--366",
    year = "1992"
}

@article{Matyjasek:2020bzc,
    author = "Matyjasek, Jerzy",
    title = "{Quasinormal modes of dirty black holes in the effective theory of gravity with a third order curvature term}",
    eprint = "2009.10793",
    archivePrefix = "arXiv",
    primaryClass = "gr-qc",
    doi = "10.1103/PhysRevD.102.124046",
    journal = "Phys. Rev. D",
    volume = "102",
    number = "12",
    pages = "124046",
    year = "2020"
}

@article{Wilczek:1977pj,
    author = "Wilczek, Frank",
    title = "{Problem of Strong  $P$  and  $T$  Invariance in the Presence of Instantons}",
    reportNumber = "Print-77-0939 (COLUMBIA)",
    doi = "10.1103/PhysRevLett.40.279",
    journal = "Phys. Rev. Lett.",
    volume = "40",
    pages = "279--282",
    year = "1978"
}

@article{Peccei:1977ur,
    author = "Peccei, R. D. and Quinn, Helen R.",
    title = "{Constraints Imposed by CP Conservation in the Presence of Instantons}",
    reportNumber = "ITP-572-STANFORD",
    doi = "10.1103/PhysRevD.16.1791",
    journal = "Phys. Rev. D",
    volume = "16",
    pages = "1791--1797",
    year = "1977"
}

@article{Metsaev:1987zx,
    author = "Metsaev, R. R. and Tseytlin, Arkady A.",
    title = "{Order alpha-prime (Two Loop) Equivalence of the String Equations of Motion and the Sigma Model Weyl Invariance Conditions: Dependence on the Dilaton and the Antisymmetric Tensor}",
    reportNumber = "PRINT-87-0184 (LEBEDEV)",
    doi = "10.1016/0550-3213(87)90077-0",
    journal = "Nucl. Phys. B",
    volume = "293",
    pages = "385--419",
    year = "1987"
}

@article{Donoghue:1993eb,
    author = "Donoghue, John F.",
    title = "{Leading quantum correction to the Newtonian potential}",
    eprint = "gr-qc/9310024",
    archivePrefix = "arXiv",
    reportNumber = "UMHEP-396",
    doi = "10.1103/PhysRevLett.72.2996",
    journal = "Phys. Rev. Lett.",
    volume = "72",
    pages = "2996--2999",
    year = "1994"
}

@article{Donoghue:2017vvl,
    author = "Donoghue, John F. and Menezes, Gabriel",
    title = "{Inducing the Einstein action in QCD-like theories}",
    eprint = "1712.04468",
    archivePrefix = "arXiv",
    primaryClass = "hep-ph",
    doi = "10.1103/PhysRevD.97.056022",
    journal = "Phys. Rev. D",
    volume = "97",
    number = "5",
    pages = "056022",
    year = "2018"
}

@article{Svrcek:2006yi,
    author = "Svrcek, Peter and Witten, Edward",
    title = "{Axions In String Theory}",
    eprint = "hep-th/0605206",
    archivePrefix = "arXiv",
    reportNumber = "SLAC-PUB-11894",
    doi = "10.1088/1126-6708/2006/06/051",
    journal = "JHEP",
    volume = "06",
    pages = "051",
    year = "2006"
}

@article{Gross:1986iv,
    author = "Gross, David J. and Witten, Edward",
    title = "{Superstring Modifications of Einstein's Equations}",
    reportNumber = "Print-86-0250 (PRINCETON)",
    doi = "10.1016/0550-3213(86)90429-3",
    journal = "Nucl. Phys. B",
    volume = "277",
    pages = "1",
    year = "1986"
}

@article{Marciu:2020ski,
    author = "Marciu, Mihai",
    title = "{Dynamical aspects for scalar fields coupled to cubic contractions of the Riemann tensor}",
    eprint = "2004.07120",
    archivePrefix = "arXiv",
    primaryClass = "gr-qc",
    doi = "10.1103/PhysRevD.102.023517",
    journal = "Phys. Rev. D",
    volume = "102",
    number = "2",
    pages = "023517",
    year = "2020"
}

@article{daRocha:2020gee,
    author = "da Rocha, Rold{\~a}o and Tomaz, Anderson A.",
    title = "{MGD-decoupled black holes, anisotropic fluids and holographic entanglement entropy}",
    eprint = "2005.02980",
    archivePrefix = "arXiv",
    primaryClass = "hep-th",
    doi = "10.1140/epjc/s10052-020-8414-8",
    journal = "Eur. Phys. J. C",
    volume = "80",
    number = "9",
    pages = "857",
    year = "2020"
}

@article{daRocha:2021sqd,
    author = "da Rocha, Roldao",
    title = "{Gravitational decoupling of generalized Horndeski hybrid stars}",
    eprint = "2111.11995",
    archivePrefix = "arXiv",
    primaryClass = "gr-qc",
    doi = "10.1140/epjc/s10052-021-09971-3",
    journal = "Eur. Phys. J. C",
    volume = "82",
    number = "1",
    pages = "34",
    year = "2022"
}

@article{Kobayashi:2019hrl,
    author = "Kobayashi, Tsutomu",
    title = "{Horndeski theory and beyond: a review}",
    eprint = "1901.07183",
    archivePrefix = "arXiv",
    primaryClass = "gr-qc",
    reportNumber = "RUP-19-3",
    doi = "10.1088/1361-6633/ab2429",
    journal = "Rept. Prog. Phys.",
    volume = "82",
    number = "8",
    pages = "086901",
    year = "2019"
}

@article{Weinberg:1977ma,
    author = "Weinberg, Steven",
    title = "{A New Light Boson?}",
    reportNumber = "HUTP-77/A074",
    doi = "10.1103/PhysRevLett.40.223",
    journal = "Phys. Rev. Lett.",
    volume = "40",
    pages = "223--226",
    year = "1978"
}

@article{Alexander:2008wi,
    author = "Alexander, Stephon and Yunes, Nicolas",
    title = "{Chern-Simons Modified Gravity as a Torsion Theory and its Interaction with Fermions}",
    eprint = "0804.1797",
    archivePrefix = "arXiv",
    primaryClass = "gr-qc",
    reportNumber = "IGC-08-4-1",
    doi = "10.1103/PhysRevD.77.124040",
    journal = "Phys. Rev. D",
    volume = "77",
    pages = "124040",
    year = "2008"
}

@article{Perrucci:2024qrr,
    author = "Perrucci, Italo and Kuipers, Folkert and Casadio, Roberto",
    title = "{Quantum gravitational hair in gravastars and observational tests}",
    eprint = "2412.04886",
    archivePrefix = "arXiv",
    primaryClass = "gr-qc",
    doi = "10.1088/1475-7516/2025/03/005",
    journal = "JCAP",
    volume = "03",
    pages = "005",
    year = "2025"
}

@article{Calmet:2019eof,
    author = "Calmet, Xavier and Casadio, Roberto and Kuipers, Folkert",
    title = "{Quantum Gravitational Corrections to a Star Metric and the Black Hole Limit}",
    eprint = "1909.13277",
    archivePrefix = "arXiv",
    primaryClass = "hep-th",
    doi = "10.1103/PhysRevD.100.086010",
    journal = "Phys. Rev. D",
    volume = "100",
    number = "8",
    pages = "086010",
    year = "2019"
}

@article{daRocha:2021xwq,
    author = "da Rocha, Roldao",
    title = "{Holographic entanglement entropy, deformed black branes, and deconfinement in AdS/QCD}",
    eprint = "2111.01244",
    archivePrefix = "arXiv",
    primaryClass = "hep-th",
    doi = "10.1103/PhysRevD.105.026014",
    journal = "Phys. Rev. D",
    volume = "105",
    number = "2",
    pages = "026014",
    year = "2022"
}

@article{daRocha:2023waq,
    author = "da Rocha, Roldao",
    title = "{Generalized extremal branes in AdS/CMT and holographic superconductors}",
    eprint = "2310.07860",
    archivePrefix = "arXiv",
    primaryClass = "hep-th",
    doi = "10.1016/j.aop.2024.169663",
    journal = "Annals Phys.",
    volume = "465",
    pages = "169663",
    year = "2024"
}

@article{tHooft:1974toh,
    author = "'t Hooft, Gerard and Veltman, M. J. G.",
    title = "{One loop divergencies in the theory of gravitation}",
    journal = "Ann. Inst. H. Poincare Phys. Theor. A",
    volume = "20",
    pages = "69--94",
    year = "1974"
}

@article{Goroff:1985th,
    author = "Goroff, Marc H. and Sagnotti, Augusto",
    title = "{The Ultraviolet Behavior of Einstein Gravity}",
    reportNumber = "CALT-68-1289, LBL-19995, UCB-PTH-85-34",
    doi = "10.1016/0550-3213(86)90193-8",
    journal = "Nucl. Phys. B",
    volume = "266",
    pages = "709--736",
    year = "1986"
}

@article{Marciu:2023hdb,
    author = "Marciu, Mihai and Ioan, Dana Maria and Dragomir, Mihai",
    title = "{Observational constraints for cubic gravity theory based on third order contractions of the Riemann tensor}",
    eprint = "2311.11297",
    archivePrefix = "arXiv",
    primaryClass = "gr-qc",
    doi = "10.1140/epjc/s10052-024-12559-2",
    journal = "Eur. Phys. J. C",
    volume = "84",
    number = "2",
    pages = "196",
    year = "2024"
}

@article{Kulkarni:2024ghc,
    author = "Kulkarni, Raghotham A. and Rahul and Bhattacharyya, Soham and Kothawala, Dawood",
    title = "{Worldline EFT treatment of quadratic and cubic gravity theories}",
    eprint = "2410.01266",
    archivePrefix = "arXiv",
    primaryClass = "gr-qc",
    doi = "10.1103/yx1z-1wzs",
    journal = "Phys. Rev. D",
    volume = "112",
    number = "12",
    pages = "124028",
    year = "2025"
}

@article{daRocha:2024lev,
    author = "da Rocha, Roldao",
    title = "{Deformations of the AdS{\textendash}Schwarzschild black brane and the shear viscosity of the quark{\textendash}gluon plasma}",
    eprint = "2409.17325",
    archivePrefix = "arXiv",
    primaryClass = "hep-th",
    doi = "10.1140/epjp/s13360-024-05795-8",
    journal = "Eur. Phys. J. Plus",
    volume = "139",
    number = "11",
    pages = "1006",
    year = "2024"
}

@article{Vacaru:2024dvc,
    author = "Vacaru, Sergiu I.",
    title = "{Asymptotic safe nonassociative quantum gravity with star R-flux products, Goroff{\textendash}Sagnotti counter-terms, and geometric flows}",
    eprint = "2410.05666",
    archivePrefix = "arXiv",
    primaryClass = "hep-th",
    doi = "10.1016/j.aop.2024.169812",
    journal = "Annals Phys.",
    volume = "470",
    pages = "169812",
    year = "2024"
}

@article{Lessa:2023thi,
    author = "Lessa, L. A. and Maluf, R. V. and Silva, J. E. G. and Almeida, C. A. S.",
    title = "{Braneworlds in warped Einsteinian cubic gravity}",
    eprint = "2312.06588",
    archivePrefix = "arXiv",
    primaryClass = "gr-qc",
    doi = "10.1088/1475-7516/2024/05/123",
    journal = "JCAP",
    volume = "05",
    pages = "123",
    year = "2024"
}

@article{DeFelice:2023vmj,
    author = "De Felice, Antonio and Tsujikawa, Shinji",
    title = "{Excluding static and spherically symmetric black holes in Einsteinian cubic gravity with unsuppressed higher-order curvature terms}",
    eprint = "2305.07217",
    archivePrefix = "arXiv",
    primaryClass = "gr-qc",
    reportNumber = "YITP-23-63, WUCG-23-06",
    doi = "10.1016/j.physletb.2023.138047",
    journal = "Phys. Lett. B",
    volume = "843",
    pages = "138047",
    year = "2023"
}

@article{Rachwal:2021bgb,
    author = "Rachwal, Leslaw and Modesto, Leonardo and Pinzul, Aleksandr and Shapiro, Ilya L.",
    title = "{Renormalization group in six-derivative quantum gravity}",
    eprint = "2104.13980",
    archivePrefix = "arXiv",
    primaryClass = "hep-th",
    doi = "10.1103/PhysRevD.104.085018",
    journal = "Phys. Rev. D",
    volume = "104",
    number = "8",
    pages = "085018",
    year = "2021"
}

@article{Battista:2023iyu,
    author = "Battista, Emmanuele",
    title = "{Quantum Schwarzschild geometry in effective field theory models of gravity}",
    eprint = "2312.00450",
    archivePrefix = "arXiv",
    primaryClass = "gr-qc",
    doi = "10.1103/PhysRevD.109.026004",
    journal = "Phys. Rev. D",
    volume = "109",
    number = "2",
    pages = "026004",
    year = "2024"
}

@article{Wang:2025fmz,
    author = "Wang, Zi-Liang and Battista, Emmanuele",
    title = "{Dynamical features and shadows of quantum Schwarzschild black hole in effective field theories of gravity}",
    eprint = "2501.14516",
    archivePrefix = "arXiv",
    primaryClass = "gr-qc",
    doi = "10.1140/epjc/s10052-025-13833-7",
    journal = "Eur. Phys. J. C",
    volume = "85",
    number = "3",
    pages = "304",
    year = "2025"
}

@article{Panotopoulos:2025ygq,
    author = "Panotopoulos, Grigoris and Tello-Ortiz, Francisco",
    title = "{Quantum black holes: Perihelion advance, quasi normal modes and classical/topological thermodynamics}",
    eprint = "2507.22945",
    archivePrefix = "arXiv",
    primaryClass = "gr-qc",
    doi = "10.1016/j.physletb.2025.139769",
    journal = "Phys. Lett. B",
    volume = "868",
    pages = "139769",
    year = "2025"
}

@article{Gonzalez-Espinoza:2026gen,
    author = "Gonzalez-Espinoza, Manuel and G{\'o}mez-Leyton, Y. and Stuchlik, Z. and Tello-Ortiz, Francisco",
    title = "{Non-Schwarzschild black holes sourced by scalar-vector fields}",
    eprint = "2603.17669",
    archivePrefix = "arXiv",
    primaryClass = "gr-qc",
    doi = "10.1103/p1b9-5hyk",
    journal = "Phys. Rev. D",
    volume = "114",
    number = "4",
    pages = "044021",
    year = "2026"
}

@article{Tello-Ortiz:2024mqg,
    author = "Tello-Ortiz, Francisco and Avalos, R. and G{\'o}mez-Leyton, Y. and Contreras, E.",
    title = "{Charged black holes by gravitational decoupling satisfying a non-local EoS}",
    doi = "10.1016/j.dark.2024.101547",
    journal = "Phys. Dark Univ.",
    volume = "46",
    pages = "101547",
    year = "2024"
}

@article{Iyer:1994ys,
    author = "Iyer, Vivek and Wald, Robert M.",
    title = "{Some properties of Noether charge and a proposal for dynamical black hole entropy}",
    eprint = "gr-qc/9403028",
    archivePrefix = "arXiv",
    doi = "10.1103/PhysRevD.50.846",
    journal = "Phys. Rev. D",
    volume = "50",
    pages = "846--864",
    year = "1994"
}

@article{Hawking:1975vcx,
    author = "Hawking, S. W.",
    editor = "Gibbons, G. W. and Hawking, S. W.",
    title = "{Particle Creation by Black Holes}",
    doi = "10.1007/BF02345020",
    journal = "Commun. Math. Phys.",
    volume = "43",
    pages = "199--220",
    year = "1975",
    note = "[Erratum: Commun.Math.Phys. 46, 206 (1976)]"
}

@article{Casadio:1997yv,
    author = "Casadio, R. and Harms, B. and Leblanc, Y.",
    title = "{Microfield dynamics of black holes}",
    eprint = "gr-qc/9712017",
    archivePrefix = "arXiv",
    reportNumber = "UAHEP-9714",
    doi = "10.1103/PhysRevD.58.044014",
    journal = "Phys. Rev. D",
    volume = "58",
    pages = "044014",
    year = "1998"
}

@article{Arnowitt:1959ah,
    author = "Arnowitt, Richard L. and Deser, Stanley and Misner, Charles W.",
    title = "{Dynamical Structure and Definition of Energy in General Relativity}",
    doi = "10.1103/PhysRev.116.1322",
    journal = "Phys. Rev.",
    volume = "116",
    pages = "1322--1330",
    year = "1959"
}
\end{document}